\documentclass{elsart}
\usepackage{epsfig}
\usepackage{pst-node}
\usepackage{pst-plot}
\usepackage{changebar}
\usepackage{setspace}

\newcounter{bla}
\newenvironment{refnummer}{%
\list{[\arabic{bla}]}%
{\usecounter{bla}%
 \setlength{\itemindent}{0pt}%
 \setlength{\topsep}{0pt}%
 \setlength{\itemsep}{0pt}%
 \setlength{\labelsep}{2pt}%
 \setlength{\listparindent}{0pt}%
 \settowidth{\labelwidth}{[9]}%
 \setlength{\leftmargin}{\labelwidth}%
 \addtolength{\leftmargin}{\labelsep}%
 \setlength{\rightmargin}{0pt}}}
 {\endlist}

\begin{document}
\begin{frontmatter}

\title{THERMUS}

\author[a]{S. Wheaton\thanksref{author}},
\author[a]{J. Cleymans},
\author[a,b]{M. Hauer}

\thanks[author]{Corresponding author: spencer.wheaton@uct.ac.za}

\address[a]{{\it UCT-CERN Research Centre and Department  of  Physics,
University of Cape Town, Rondebosch 7701, South Africa}}
\address[b]{Helmholtz Research School, University of Frankfurt, Frankfurt, Germany}

\begin{abstract}

THERMUS is a package of C++ classes and functions allowing statistical-thermal 
model analyses of particle production in relativistic heavy-ion collisions to 
be performed within the ROOT framework of analysis. Calculations are possible 
within three statistical ensembles; a grand-canonical treatment of the 
conserved charges $B$, $S$ and $Q$, a fully canonical treatment of the 
conserved charges, and a mixed-canonical ensemble combining a canonical 
treatment of strangeness with a grand-canonical treatment of baryon number and 
electric charge. THERMUS allows 
for the 
assignment of decay chains and detector efficiencies specific to each 
particle yield, which enables sensible fitting of model parameters to 
experimental data.    

\begin{flushleft}

\end{flushleft}


\end{abstract}

\end{frontmatter}


\pagebreak

{\bf PROGRAM SUMMARY}

\begin{small}
\noindent
{\em Manuscript Title:} THERMUS - A Thermal Model Package for ROOT \\
{\em Authors:} S. Wheaton, J. Cleymans, M. Hauer \\
{\em Program Title:} THERMUS, version 2.1 \\
{\em Journal Reference:}                                      \\
{\em Catalogue identifier:}                                   \\
{\em Licensing provisions:} none                                  \\
{\em Programming language:} C++                                   \\
{\em Computer:} PC, Pentium 4, 1 GB RAM (not hardware dependent)                         \\
{\em Operating system:} Linux: FEDORA, RedHat etc                                        \\
{\em RAM:}                                               
{\em Classification:} 17.7 Experimental Analysis - Fission, Fusion, Heavy-ion                                         \\
{\em External routines/libraries:} 'Numerical Recipes in C' [1], ROOT [2]                                     \\

{\em Nature of problem:}\\

Statistical-thermal model analyses of heavy-ion collision data require the 
calculation of both primordial particle densities and contributions from 
resonance decay. A set of thermal parameters 
(the number depending on the particular model imposed) and a set of thermalised 
particles, with their decays specified, is required as input to these models. The 
output is then a complete set of primordial thermal quantities for each particle, 
together with the contributions to the final particle yields from resonance decay.\\ 

In many applications of statistical-thermal models it is required to fit experimental 
particle multiplicities or particle ratios. In such analyses, the 
input is a set of experimental yields and ratios, a set of particles comprising 
the assumed hadron resonance gas formed in the collision and the constraints to be placed on the system. The thermal model parameters 
consistent with the specified constraints leading to the best-fit 
to the experimental data are then output.\\    

{\em Solution method:}\\

THERMUS is a package designed for incorporation into the ROOT [2] framework, used extensively by the 
heavy-ion community. As such, it utilises a great deal of ROOT's functionality in its operation. Three 
distinct statistical ensembles are included in THERMUS, while additional options to include 
quantum statistics, resonance width and excluded volume corrections are also available.\\

THERMUS provides a default particle list including all mesons (up to the $K_4^*(2045)$) and 
baryons (up to the $\Omega^-$) listed in the July 2002 Particle Physics Booklet [3]. 
For each typically unstable particle in 
this list, THERMUS includes a text-file listing its decays. With thermal parameters specified, 
THERMUS calculates primordial thermal densities either by 
performing numerical integrations or else, in the case of the Boltzmann approximation without resonance width in the grand-canonical ensemble, 
by evaluating Bessel functions. Particle decay chains are then used to evaluate 
experimental observables (i.e. particle yields following resonance decay). Additional detector efficiency 
factors allow fine-tuning of the model predictions to a specific detector arrangement.\\

When parameters are required to be constrained, use is made of the `Numerical Recipes in C' [1] function 
which applies the Broyden globally
convergent secant method of solving nonlinear systems of equations. THERMUS provides the means 
of imposing a large number of constraints on the chosen model (amongst others, THERMUS can fix the 
baryon-to-charge ratio of the system, the strangeness density of the system and the primordial energy 
per hadron).\\

Fits to experimental data are accomplished in THERMUS by using the ROOT \texttt{TMinuit} class. 
In its default operation, the standard $\chi^2$ function is minimised, yielding the set of best-fit thermal 
parameters. THERMUS allows the assignment of separate decay chains to each experimental input. 
In this way, the model is able to match the specific feed-down corrections of a particular data 
set.\\ 

{\em Running time:} Depending on the analysis required, run-times vary from seconds 
(for the evaluation of particle multiplicities given a set of parameters) to several minutes 
(for fits to experimental data subject to constraints).\\
   \\
{\em References:}
\begin{refnummer}
\item  W. H. Press, S. A. Teukolsky, W. T. Vetterling, B. P. Flannery, Numerical Recipes in C: The Art of Scientific Computing (Cambridge University Press, Cambridge, 2002).
\item  R. Brun and F. Rademakers, Nucl. Inst. \& Meth. in Phys. Res. A {\bf 389} (1997) 81.\\ 
  See also http://root.cern.ch/.         
\item K. Hagiwara \textit{et al.}, Phys. Rev. D {\bf 66} (2002) 010001.
\end{refnummer}

\end{small}

\newpage


\section{Introduction}

The statistical-thermal model has proved extremely successful~\cite{andronic,wheaton,becattini} in 
describing the hadron multiplicities 
observed in relativistic collisions of both heavy-ions and elementary particles.
The methods used in calculating these yields have been extensively reviewed in recent 
years~\cite{jaipur,reviewPBM}.  The success of these models has led to the creation 
of several software codes~\cite{share,sharev2,therminator} that use 
experimental particle yields as input and calculate the corresponding chemical
freeze-out temperature ($T$)  and baryon chemical potential ($\mu_B$).
In this paper we present THERMUS, a package of C++ classes 
and functions, which is based on 
the object-oriented  ROOT framework~\cite{Root}.  All THERMUS C++ classes inherit from the ROOT base class \texttt{TObject}. This
allows them to be fully integrated into the interactive ROOT environment, allowing all of the 
ROOT functionality in a statistical-thermal model analysis. Recent applications of 
THERMUS include~\cite{wheaton,LHC,Kraus,Caines,Stiles,Takahashi,Murray,HauerViscosity,Witt,Hippo,Sahoo}. 
An on-going effort to extend the range of applications of THERMUS has led to several publications on 
fluctuations in statistical models~\cite{Hauer1,Hauer2,Hauer3}.\\

The paper is structured in the following way. In Section 2 an overview is presented of the theoretical model on which 
THERMUS is based. Section 3 outlines the structure and functionality of the THERMUS code, while Section 4 explains 
the installation procedure.\\

\section{Overview of the Statistical-Thermal Model of Heavy-Ion Collisions}

\subsection{Choice of Ensemble}
                
Within the statistical-thermal model there is a freedom regarding the ensemble
with which to treat the quantum numbers $B$ (baryon number), $S$ (strangeness) 
and $Q$ (charge), which are
conserved in strong interactions. The introduction of chemical 
potentials for each of these quantum numbers 
(i.e.\ a grand-canonical description) allows fluctuations about
conserved averages. This is a reasonable approximation only when the
number of particles carrying the quantum number concerned is large. In
applications of the thermal model to high-energy elementary
collisions, such as $pp$, $p\bar{p}$ and $e^+e^-$ collisions~\cite{B1,B2}, a
canonical treatment of each of the quantum numbers is required. Within
such a canonical description, quantum numbers are conserved
exactly. In small systems or at low temperatures (more specifically,
low $VT^3$ values), a canonical treatment leads to a suppression of
hadrons carrying non-zero quantum numbers, since these particles have
to be created in pairs. In heavy-ion collisions, the large number of
baryons and charged particles generally allows baryon number and charge to be
treated grand-canonically. However, at the low temperatures of the GSI
SIS, the resulting low production of strange particles requires a
canonical treatment of strangeness~\cite{C8}. This is the so-called
mixed-canonical approach. \\ 

In order to calculate the thermal properties of a system, one starts with 
an evaluation of its partition function. The form of the partition function 
obviously depends on the choice of ensemble. In the following sections, we consider 
the three ensembles most widely used in applications of the statistical-thermal model.\\

\subsubsection{The Grand-Canonical Ensemble}\label{SubSection::GCanEnsemble}

This ensemble is the most widely used in applications to heavy-ion 
collisions~\cite{reviewPBM,review,heavy_ions,PBM_qd,abundances_a,PBMRHIC,abundances_b,Polish_a,Polish_b,Xu,Kaneta}. 
Within this ensemble, conservation laws for energy and quantum or particle numbers 
are enforced on average through the temperature and chemical potentials.\\

In the case of a multi-component hadron gas of volume $V$ and temperature $T$, the 
logarithm of the total partition 
function is given by,
\begin{eqnarray}
\ln Z^{GC}(T,V,\{\mu_i\}) &=& \sum_{\mathrm{species}\;i}{g_iV\over\left(2\pi\right)^3}\int d^3p\ln\left(1\pm e^{-\beta\left(E_i-\mu_i\right)}\right)^{\pm1},
                          \end{eqnarray}
where $g_i$ and $\mu_i$ are, respectively, the degeneracy and chemical potential 
of hadron species $i$, $\beta\equiv 1/T$, while $E_i=\sqrt{p^2+m_i^2}$, where $m_i$ is the particle 
mass. The plus sign refers to fermions and the minus sign to bosons.

Since in relativistic heavy-ion collisions it is not individual particle numbers 
that are conserved, but rather the quantum numbers $B$, $S$ and $Q$, the chemical 
potential for particle species $i$ is given by,
\begin{eqnarray}
\mu_i &=& B_i\mu_B + S_i\mu_S + Q_i\mu_Q,
\end{eqnarray}
where $B_i$, $S_i$ and $Q_i$ are the baryon number, strangeness and charge, respectively, of hadron species 
$i$, and $\mu_B$, $\mu_S$ and $\mu_Q$ are the corresponding chemical potentials for 
these conserved quantum numbers.\\

Once the partition function is known, the particle 
multiplicities, entropy and pressure are obtained by differentiation: 
\begin{eqnarray}
N_i^{GC} &=& T{\partial\ln Z^{GC}\over\partial\mu_i},\\
S^{GC} &=& {\partial\over\partial T}\left(T\ln Z^{GC}\right),\\ 
P^{GC} &=& T{\partial\ln Z^{GC}\over\partial V}.
\end{eqnarray}
Furthermore, the energy is given by,
\begin{eqnarray}
E^{GC} &=& T^2\frac{\partial \ln Z^{GC}}{\partial T} + \sum_{\mathrm{species}\;i}\mu_i\;N_i^{GC}.
\end{eqnarray}
Using the prescription for the particle multiplicity, 
\begin{eqnarray}
N_i^{GC} &=& {g_iV \over 2\pi^2}\sum_{k=1}^{\infty}(\mp
  1)^{k+1}{m_i^2 T\over k}
  K_2\left(km_i\over T\right)e^{\beta k\mu_i}=\sum_{k=1}^{\infty}z_i^k e^{\beta k \mu_i},\label{density}
\end{eqnarray}
where we have introduced the $z_i^k$. Similar expressions exist for the energy, entropy and pressure.\\

In practice, the Boltzmann approximation (i.e. retaining just the 
$k=1$ term in Equation~(\ref{density})) is reasonable 
for all particles except the pions. In this approximation,
\begin{eqnarray}
\ln Z^{GC}\left(T,V,\{\mu_i\}\right) &=& \sum_{\mathrm{species}\;i}\frac{g_iV}{\left(2\pi\right)^3}\int d^3p\;e^{-\beta\left(E_i-\mu_i\right)}= \sum_{\mathrm{species}\;i} z_i^1\;e^{\beta\mu_i},
\end{eqnarray}
where $z_i^1$ is the single-particle partition function of hadron species $i$. 
Furthermore, under this approximation, 
$P=\sum_{\mathrm{species}\;i}N_i^{GC}T/V$ both for massive and massless particles, which is certainly not 
true for quantum statistics.\\ 

Since the use of quantum statistics requires numerical integration (or evaluation of 
infinite sums), while Boltzmann statistics can be implemented analytically, it 
is worthwhile to identify those regions in which quantum statistics deviate 
greatly from Boltzmann statistics. In most applications of the 
statistical-thermal model, only a small region of the $\mu-T$ parameter space is of 
interest. Using the freeze-out condition of constant $E/N$~\cite{UnifiedFO}, the 
thermal parameters, and hence the percentage deviation from Boltzmann statistics, can be determined 
as a function of the collision energy $\sqrt{s}$. From such an analysis it is evident 
that, for pions, quantum statistics must be implemented at all 
but the lowest energies (deviation at the level of 10\%), while, for kaons, 
the deviation peaks at between 1 and 2\%. For all other mesons, the deviation is below the 1\% level. 
For baryons, the deviation is extremely small for all except the protons at small $\sqrt{s}$.\\

When quantum statistics are applied, restrictions have to be imposed on the chemical 
potentials so as to avoid Bose-Einstein condensation. The Bose-Einstein distribution 
function diverges if,
\begin{eqnarray}
e^{\beta\left(m_i-\mu_i\right)} \le 1.
\end{eqnarray}
Such Bose-Einstein condensation is avoided, provided that the 
chemical potentials of all bosons included in the resonance gas are 
less than their masses (i.e. $\mu_i<m_i$).\\  

\subsubsection{The Canonical Ensemble}\label{SubSection::Canonical}

Within this ensemble, quantum number conservation is exactly enforced. Considering the fully canonical treatment of $B$, $S$ 
and $Q$ in the Boltzmann approximation, as investigated 
in~\cite{Keranen1}, the partition function for the system is given by,
\begin{eqnarray}
Z_{B,S,Q} &=& {Z_0\over(2\pi)^2}\int_{-\pi}^{\pi}d\phi_S\;
\int_{-\pi}^{\pi}d\phi_Q\;\cos\left(S\phi_S + Q\phi_Q  - B\arg\omega\right)\nonumber\\
& &\times\exp\left[2\sum_{\mathrm{mesons}\;j}z_j^1\;\cos\left(S_j\phi_S+Q_j\phi_Q\right)\right]
I_B\left(2|\omega|\right),\label{Eq:BSQ Canonical}
\end{eqnarray}
where,
\begin{eqnarray}
\omega &\equiv& \sum_{\mathrm{baryons}\;j}z_j^1\;e^{i(S_j\phi_S+Q_j\phi_Q)},\nonumber\\
z_j^1 &\equiv& {g_jV\over\left(2\pi\right)^3}\int d^3p\;e^{-\beta E_j},\nonumber
\end{eqnarray}
$Z_0$ represents the contribution of those hadrons with no net charges, and the sums over mesons 
and baryons extend only over the particles (i.e. not the anti-particles).\\

Once the partition function is known, we can calculate all
thermodynamic properties of the system. Using thermodynamic relations it follows that,
\begin{eqnarray}
S &=&{\partial \over\partial T}\left(T\ln Z_{B,S,Q}\right),
\end{eqnarray}
and,
\begin{eqnarray}
P &=&T{\partial\ln Z_{B,S,Q} \over\partial V}.
\end{eqnarray}

Furthermore, the multiplicity of hadron species $i$ within this ensemble, $N_i^{B,S,Q}$, is calculated by 
multiplying the single-particle partition function for particle $i$, appearing in the canonical 
partition function, by a fictitious fugacity $\lambda_i$, differentiating with respect 
to $\lambda_i$, and then setting $\lambda_i$ to 1:
\begin{eqnarray}
  N_i^{B,S,Q} &=& \left.{\partial\ln
  Z_{B,S,Q}({\lambda_i})\over\partial\lambda_i}\right|_{\lambda_i=1}.
\end{eqnarray}

Following these prescriptions,\\
\begin{eqnarray}
  N_i^{B,S,Q} &=& \left({Z_{B-B_i,S-S_i,Q-Q_i}\over
  Z_{B,S,Q}}\right)\left.N_i^{GC}\right|_{\mu_i=0},
\end{eqnarray}
\begin{eqnarray}
  S^{B,S,Q} &=& \ln Z_{B,S,Q} + \sum_{\mathrm{species}\;i} \left({Z_{B-B_i,S-S_i,Q-Q_i}\over
    Z_{B,S,Q}}\right){\left.E_i^{GC}\right|_{\mu_i=0}\over T},\\
  P^{B,S,Q} &=&  \sum_{\mathrm{species}\;i} \left({Z_{B-B_i,S-S_i,Q-Q_i}\over
    Z_{B,S,Q}}\right)\left.P_i^{GC}\right|_{\mu_i=0},\\
  E^{B,S,Q} &=& \sum_{\mathrm{species}\;i} \left({Z_{B-B_i,S-S_i,Q-Q_i}\over
    Z_{B,S,Q}}\right)\left.E_i^{GC}\right|_{\mu_i=0}.
\end{eqnarray}

One notices that, in the Boltzmann approximation, the
particle and energy density and pressure of particle species $i$, within the
canonical ensemble, differ from that in the grand-canonical formalism, with all chemical 
potentials set to zero, by a multiplicative factor $\left({Z_{B-B_i,S-S_i,Q-Q_i}/Z_{B,S,Q}}\right)$. 
This correction factor depends only on the thermal parameters of the 
system and the quantum numbers of the particle (i.e. the correction for the 
$\Delta^+$ and $p$ are the same). The
entropy is, however, slightly different; the total entropy cannot be
split into the sum of contributions from separate particles.\\

Now, 
\begin{eqnarray}
  \lim_{V\rightarrow\infty}\left({Z_{B-B_i,S-S_i,Q-Q_i}\over
    Z_{B,S,Q}}\right) = e^{B_i\mu_B/T}e^{S_i\mu_S/T}e^{Q_i\mu_Q/T}.\label{Eq:Corr Factors}
  \end{eqnarray}
Thus, for large systems, the grand-canonical results for the particle number, entropy, 
pressure and energy are approached~\cite{Keranen1}.\\

\subsubsection{The Mixed-Canonical (Strangeness-Canonical) Ensemble}\label{SubSection::SCanonical}

Within this ensemble, the strangeness in the system is fixed exactly by its 
initial value of $S$, while the baryon and charge content are treated grand-canonically. 
For a Boltzmann hadron gas of strangeness $S$,
\begin{eqnarray}
Z_{S} &=& {1\over2\pi}\int_{-\pi}^{\pi}d\phi_S\;e^{-iS\phi_S}\nonumber\\
& & \times\exp\left[\sum_{\mathrm{hadrons}\;i}{g_iV\over\left(2\pi\right)^3}\int
d^3p\;e^{-\beta\left(
  E_i-\mu_i\right)}\;e^{iS_i\phi_S}\right],
\end{eqnarray}
where the sum over hadrons includes both particles and
anti-particles and,
\begin{eqnarray}
\mu_i &=& B_i\mu_B + Q_i\mu_Q.
\end{eqnarray}

Applying the same prescription for the evaluation of the particle multiplicities as discussed 
for the canonical ensemble, it follows that,
\begin{eqnarray}
  N_i^{S} &=& \left({Z_{S-S_i}\over
  Z_{S}}\right)\left.N_i^{GC}\right|_{\mu_S=0}.
\end{eqnarray}
Furthermore,
\begin{eqnarray}
  S^{S} &=& \ln Z_{S} + \sum_{\mathrm{species}\;i} \left({Z_{S-S_i}\over
    Z_{S}}\right)\left({\left.E_i^{GC}\right|_{\mu_S=0}-\mu_i\left.N_i^{GC}\right|_{\mu_S=0}\over T}\right),\\
  P^{S} &=&  \sum_{\mathrm{species}\;i} \left({Z_{S-S_i}\over
    Z_{S}}\right)\left.P_i^{GC}\right|_{\mu_S=0},\\
  E^{S} &=& \sum_{\mathrm{species}\;i} \left({Z_{S-S_i}\over
    Z_{S}}\right)\left.E_i^{GC}\right|_{\mu_S=0}.
\end{eqnarray}

As in the case of the canonical ensemble, the strangeness-canonical results, in the Boltzmann 
approximation, differ from those 
of the grand-canonical ensemble, with $\mu_S=0$, by multiplicative correction factors which 
depend, in this case, only on the thermal parameters and the strangeness of the particle concerned. For 
large systems and high temperatures, these correction 
factors approach the grand-canonical fugacities, i.e.,
\begin{equation}
\lim_{V\rightarrow \infty}\left({Z_{S-S_i}\over Z_{S}}\right)= e^{S_i\mu_S/T}.
\end{equation}
The expression for $Z_{S}$ can be reduced~\cite{PBMCleymans} to,
\begin{eqnarray}
Z_{S} = Z_0\times\sum_{m=-\infty}^{+\infty}\sum_{n=-\infty}^{+\infty} I_{|3m+2n-S|}(x_1)&&\;I_{|n|}(x_2)\;I_{|m|}(x_3)\\\nonumber
&&\times(y_3/y_1^3)^m\;(y_2/y_1^2)^n\;y_1^{S},
\end{eqnarray}
where $Z_0$ is the contribution to the total partition function 
of the non-strange hadrons, while,
\begin{eqnarray}
x_i &=& 2\;\sqrt{k_{+i}k_{-i}}\hspace{1cm}(i=1,2,3),
\end{eqnarray}
and,
\begin{eqnarray}
y_i &=& \sqrt{{k_{+i}\over k_{-i}}}\hspace{1.8cm}(i=1,2,3),
\end{eqnarray}
with,
\begin{eqnarray}
k_{m} = \sum_{\mathrm{hadrons}\;j\;\mathrm{with}\;S_j=m}\left.n_j^{GC}\right|_{\mu_S=0}V.
\end{eqnarray}

In~\cite{RedlichTounsi,CleymansSIS} it is suggested that two volume parameters be used 
within canonical ensembles; the fireball volume at freeze-out, $V_f$, which 
provides the overall normalisation factor fixing the particle multiplicities from 
the corresponding densities, and the correlation volume, $V_c$, within which 
particles fulfill the requirement of local conservation of quantum numbers. In 
this way, by taking $V_c<V_f$, it is possible to boost the strangeness 
suppression. In fact, this was shown to be required to reproduce experimental 
heavy-ion collision data~\cite{RedlichTounsi,CleymansSIS}.\\ 

\subsection{Additional Features}

\subsubsection{Feeding from Unstable Particles}

\begin{figure}
\begin{center}
\includegraphics[width=12cm]{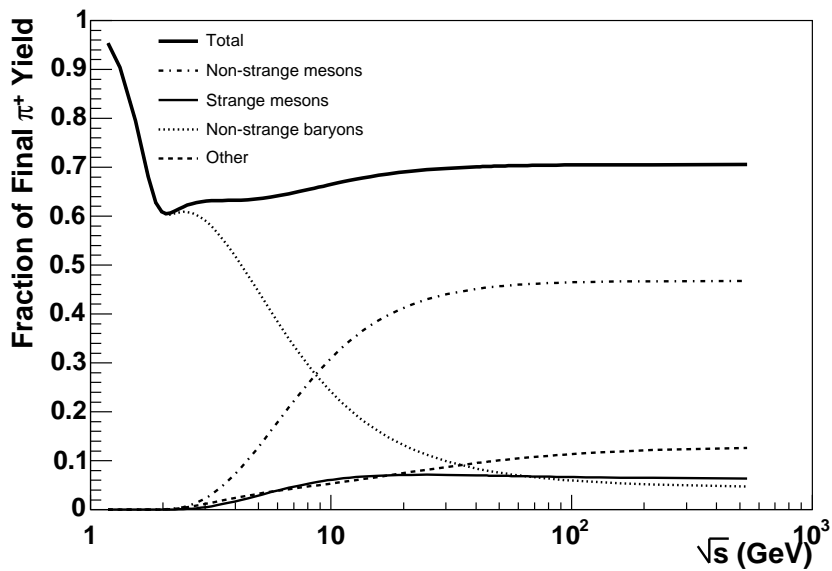}
\end{center}
\caption{The energy dependence of the decay contribution to the final $\pi^+$ yield as predicted by the 
statistical-thermal model. Calculations performed within the grand-canonical formalism, 
assuming the constant $E/N$ freeze-out criterion~\cite{UnifiedFO} (weak decays excluded).}\label{PiDecay}
\end{figure}

Since the particle yields measured by the detectors in collision experiments include 
feed-down from heavier hadrons and hadronic resonances, the primordial
hadrons are allowed to decay to particles considered stable by the
experiment before model predictions are compared with experimental
data. For example, the total $\pi^+$ yield is given by,
\begin{equation}
N_{\pi^+} = \sum_{\mathrm{species}\;i}N_i^{(\mathrm{prim})}\:Br(i\rightarrow \pi^+), 
\end{equation}
where $Br(i\rightarrow \pi^+)$ is the number of $\pi^+$'s into which a single 
particle of species $i$ decays. As shown in Figure~\ref{PiDecay}, approximately 
70\% of $\pi^+$'s originate from resonance decay at RHIC energies. Thus, a full 
treatment of resonances is essential in any statistical-thermal analysis.\\


 
\subsubsection[Resonances and the Inclusion of Resonance
        Width]{Resonances and the Inclusion of Resonance\\
        Width}\label{SubSection::ResonanceWidth}

The inclusion of a mass cut-off in the measured resonance mass
spectrum is motivated by the realisation that the time scale of a
relativistic collision does not allow the heavier resonances to reach
chemical equilibrium~\cite{SKH}. This assumes that inelastic
collisions drive the system to chemical equilibrium. If the
hadronisation process follows a statistical rule, then all resonances
should, in principle, be included~\cite{abundances_b}. This is problematic,
since data on the heavy resonances is sketchy. The situation is saved
by the finite energy density of the system, resulting in a chemical freeze-out
temperature at RHIC of approximately 160-170~MeV~\cite{Kaneta1,WheatonKaneta}, which strongly suppresses
these heavy resonances and justifies their exclusion from the
model. It is, however, important to check the sensitivity of the
extracted thermal parameters to the chosen cut-off.\\ 

The finite width of the resonances is especially important at the low
temperatures of the SIS. Resonance widths are included in the thermal
model by distributing the resonance masses according to Breit-Wigner
forms~\cite{B1,B2,heavy_ions,C8,SKH,BGS}. This amounts to the following
modification in the integration of the Boltzmann factor~\cite{C8}:  
\begin{eqnarray}
\lefteqn{\int d^3p\; \exp\left[-{\sqrt{p^2+m^2}\over T}\right]} \nonumber \\    
& & \rightarrow \int d^3p \int ds\; \exp\left[-{\sqrt{p^2+s}\over T}\right]{1 \over \pi}{m\Gamma \over (s-m^2)^2+m^2\Gamma^2}, 
\end{eqnarray}
where $\Gamma$ is the width of the resonance concerned, with threshold
limit $m_{\mathrm{threshold}}$ and mass $m$, and $\sqrt{s}$ is
integrated over the interval [$m$ - $\delta m$, $m$ + $2\Gamma$], where
\linebreak $\delta m$ = min[$m$ - $m_{\mathrm{threshold}}$,
$2\Gamma$].\\ 

\subsubsection{Deviations from Equilibrium Levels}\label{SubSection::Gammas}

The statistical-thermal model applied to elementary $e^+e^-$, $pp$ and
$p\bar{p}$ collisions~\cite{B1,B2} indicates the need for an
additional parameter, $\gamma_S$ (first introduced as a purely 
phenomenological parameter~\cite{Raf_a,Raf_b}), to account for the
observed deviation from chemical equilibrium in the strange
sector. Since a canonical ensemble was considered in these analyses,
there is an additional strangeness suppression at work, on top of the
canonical suppression. Although strangeness production is expected to
be greatly increased in $AA$ collisions, due to the larger interaction
region and increased hadron rescattering, a number of recent analyses 
\cite{BGS,BecEnergyScan,previous_analyzes,PRC,Nantes,SQMPeter} have found such a factor necessary to
accomplish a satisfactory description of data.\\

Allowance for possibly incomplete strangeness equilibration is made by 
multiplying the Boltzmann factors of each particle species in the 
partition function (or thermal distribution function $f_i(x,p)$) 
by $\gamma_S^{\left|S_i\right|}$, where $\left|S_i\right|$ is 
the number of valence strange quarks and anti-quarks in species 
$i$. The value $\gamma_S = 1$ obviously corresponds to
complete strangeness equilibration.\\  

It has been suggested~\cite{Rafelski99} that a similar parameter, 
$\gamma_q$, should be included in thermal analyses to allow for 
deviations from equilibrium levels in the non-strange sector. 
Furthermore, as collider energies increase, so does the need for the inclusion of 
charmed particles in the statistical-thermal model, with their 
occupation of phase-space possibly governed by an additional 
parameter, $\gamma_C$.\\  

\subsubsection{Excluded Volume Corrections (Grand-Canonical Ensemble)}\label{Excl Vol Corrections}

At very high energies, the ideal gas assumption is inadequate. In fact, the 
total particle densities predicted by the thermal model, with parameters 
extracted from fits to experimental data, far exceed reasonable estimates 
and measurements based on yields and the system size inferred by pion 
interferometry~\cite{VDW3}. It becomes necessary to take into account the Van der Waals--type excluded 
volume procedure~\cite{VDW3,VDW1,VDW2}. At the same fixed $T$ and $\mu_B$, all 
thermodynamic functions of the hadron gas are smaller than in the ideal 
hadron gas, and strongly decrease with increasing excluded volume.\\

Van der Waals--type corrections are included by making the following substitution for the volume $V$ in 
the canonical (with respect to particle number) partition function, 
\begin{eqnarray}
V &\rightarrow& V-\sum_{\mathrm{hadrons}\;i}\nu_i N_i, 
\end{eqnarray}
where $N_i$ is the number of hadron species $i$, and 
$\nu_i=4\left(4/3\pi r_i^3\right)$ is its proper volume, with $r_i$ its 
hard-sphere radius. This then leads to the following transcendental equation for the 
pressure of the gas in the grand-canonical ensemble (assuming $h$ particle species):
\begin{eqnarray}
P(T,\mu_1,...,\mu_h) &=& \sum_{i=1}^{h}P_i^{ideal}(T,\tilde{\mu_i}),
\end{eqnarray}
with,
\begin{eqnarray}
\tilde{\mu_i} &=& \mu_i - \nu_i P(T,\mu_1,...,\mu_h).
\end{eqnarray}
The particle, entropy and energy densities are given by,
\begin{eqnarray}
n_i(T,\mu_1,...,\mu_h) &=& \frac{n_i^{ideal}(T,\tilde{\mu_i})}{1+\sum_j\nu_jn_j^{ideal}(T,\tilde{\mu_j})},\\
s(T,\mu_1,...,\mu_h) &=& \frac{\sum_i s_i^{ideal}(T,\tilde{\mu_i})}{1+\sum_j\nu_jn_j^{ideal}(T,\tilde{\mu_j})},
\end{eqnarray}
and,
\begin{eqnarray}
e(T,\mu_1,...,\mu_h) &=& \frac{\sum_i e_i^{ideal}(T,\tilde{\mu_i})}{1+\sum_j\nu_jn_j^{ideal}(T,\tilde{\mu_j})},
\end{eqnarray}
respectively. One sees that two suppression factors enter. The first suppression is due to the shift in chemical 
potential $\mu_i\rightarrow\tilde{\mu_i}$. In the Boltzmann approximation, this leads to a 
suppression factor $e^{-\nu_iP/T}$ in all thermodynamic quantities. The second suppression is due to the 
$[1+\sum_j\nu_jn_j^{ideal}(T,\tilde{\mu_j})]^{-1}$ factor.\\

In ratios of particle numbers, although the denominator correction cancels 
out, the shift in chemical potentials leads to a change in the case 
of quantum statistics. In the Boltzmann case, even these corrections cancel out, 
provided that the same proper volume parameter $\nu$ is applied to all species.\\

\section{The Structure of THERMUS}

\subsection{Introduction}

The three distinct ensemble choices outlined in 
Sections~\ref{SubSection::GCanEnsemble}-\ref{SubSection::SCanonical} are 
implemented in THERMUS. As input to the various thermal model formalisms 
one needs first a set of particles to be considered 
thermalised. When combined with a set of thermal parameters, all primordial 
densities (i.e. number density as well as energy and entropy density and pressure) are 
calculable. Once the particle decays are known, sensible comparisons can be made with 
experimentally measured yields.\\

In THERMUS, the following units are used for the parameters:\\

\begin{center}
\begin{tabular}{lc}\hline
\multicolumn{1}{c}{Parameter} & Unit \\\hline\hline
Temperature ($T$)      & GeV \\
Chemical Potential ($\mu$) & GeV \\
Radius & fm  \\\hline
\end{tabular}
\end{center}

\noindent
Quantities frequently output by THERMUS are in the following units:\\ 

\begin{center}
\begin{tabular}{lc}\hline
\multicolumn{1}{c}{Quantity} & Unit \\\hline\hline
Number Densities ($n$) & $\mathrm{fm}^{-3}$ \\
Energy Density ($e$) & $\mathrm{GeV.fm}^{-3}$ \\
Entropy Density ($s$) & $\mathrm{fm}^{-3}$ \\
Pressure ($P$) & $\mathrm{GeV.fm}^{-3}$ \\
Volume ($V$) & $\mathrm{fm}^3$ \\\hline
\end{tabular}
\end{center}

\begin{figure}
\begin{center}
\includegraphics[width=12cm]{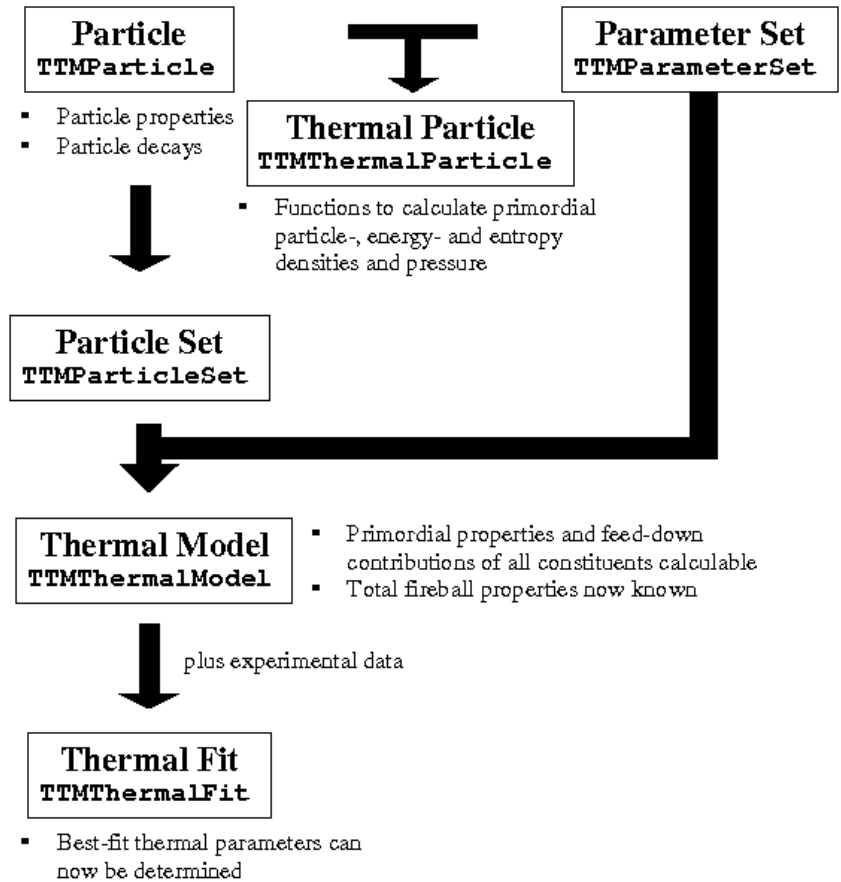}
\caption{The basic structure of THERMUS (only the most fundamental base classes are shown).}\label{FlowChart}
\end{center}
\end{figure}

In the subsections to follow, we explain the basic structure and functionality of THERMUS (shown diagrammatically in Figure~\ref{FlowChart}) by introducing 
the major THERMUS classes in a bottom-up approach. We begin with a look at the \texttt{TTMParticle} object.\footnote{It is a requirement that all ROOT 
classnames begin with a `T'. THERMUS classnames begin with `TTM' for easy identification.}\\ 

\pagebreak

\subsection{The \texttt{TTMParticle} Class} 

The properties of a particle applicable to the statistical-thermal model are grouped 
in the basic \texttt{TTMParticle} object:

\small
\begin{verbatim}


       ********* LISTING FOR PARTICLE Delta(1600)0 *********


                 ID = 32114

                 Deg. = 4

                 STAT = 1

                 Mass           =    1.6 GeV
                 Width          =   0.35 GeV
                 Threshold      = 1.07454 GeV

                 Hard sphere radius = 0

                 B = 1
                 S = 0                  |S| = 0
                 Q = 0
                 Charm = 0              |C| = 0
                 Beauty = 0
                 Top = 0

                 UNSTABLE

                 Decay Channels:


                 Summary of Decays:
         **********************************************
\end{verbatim}
\normalsize

Besides the particle name, `Delta(1600)0' in this case, its Monte Carlo numerical ID is also 
stored. This provides a far more convenient means of referencing the 
particle. The particle's decay status is also noted. In this case, the 
$\Delta(1600)^0$ is considered unstable. Particle properties are input 
using the appropriate `setters'.\\

\subsubsection{Inputting and Accessing Particle Decays}
The \texttt{TTMParticle} class allows also for the storage of a particle's 
decays. These can be entered 
from file. As an example, consider the decay file of the $\Delta(1600)^0$:

\small
\begin{verbatim}

11.67   2112    111
5.83    2212    -211
29.33   2214    -211
3.67    2114    111
22.     1114    211
8.33    2112    113
4.17    2212    -213
15.     12112   111
7.5     12212   -211

\end{verbatim}
\normalsize

Each line in the decay file corresponds to a decay channel.
The first column lists the branching ratio of the channel, while the 
subsequent tab-separated integers represent the Monte Carlo ID's of 
the daughters (each line (channel) can contain any number of daughters). 
The decay channel list of a \texttt{TTMParticle} object is populated 
with \texttt{TTMDecayChannel} objects by the 
\texttt{SetDecayChannels} function, 
with the decay file the first argument (only that part of the output that 
differs from the previous listing of the particle information is shown): 

\small
\begin{verbatim}

root [ ] part->SetDecayChannels("$THERMUS/particles/Delta\(1600\)0_decay.txt")
root [ ] part->List()

         ********* LISTING FOR PARTICLE Delta(1600)0 *********

                                     -
                                     -
                                     -


                 UNSTABLE

                 Decay Channels:
                 BRatio: 0.1167         Daughters:      2112    111
                 BRatio: 0.0583         Daughters:      2212    -211
                 BRatio: 0.2933         Daughters:      2214    -211
                 BRatio: 0.0367         Daughters:      2114    111
                 BRatio: 0.22           Daughters:      1114    211
                 BRatio: 0.0833         Daughters:      2112    113
                 BRatio: 0.0417         Daughters:      2212    -213
                 BRatio: 0.15           Daughters:      12112   111
                 BRatio: 0.075          Daughters:      12212   -211




                 Summary of Decays:
                2112            20%
                111             30.34%
                2212            10%
                -211            42.66%
                2214            29.33%
                2114            3.67%
                1114            22%
                211             22%
                113             8.33%
                -213            4.17%
                12112           15%
                12212           7.5%
         **********************************************

\end{verbatim}
\normalsize

In many cases, the branching 
ratios of unstable hadrons do not sum to 100\%. This can, however, be enforced by 
scaling all branching ratios. This is achieved when the second argument of 
\texttt{SetDecayChannels} is set to true (it is false by default).\\

In addition to the list of decay channels, a summary list of 
\texttt{TTMDecay} objects is generated in which each daughter 
appears only once, together with its total
decay fraction. This summary list is automatically generated from the
decay channel list when the \texttt{SetDecayChannels} function is called.\\
 
An existing \texttt{TList} can be set as the decay channel list of the particle, 
using the \texttt{SetDecayChannels} function. This function calls
\texttt{UpdateDecaySummary}, thereby automatically ensuring consistency between
the decay channel and decay summary lists.\\

The function \texttt{SetDecayChannelEfficiency} sets 
the reconstruction efficiency of the specified
decay channel to the specified percentage. Again, a consistent decay summary list 
is generated.\\
 
Access to the \texttt{TTMDecayChannel} objects in the decay channel list is achieved 
through the \texttt{GetDecayChannel} method. If the extracted decay channel 
is subsequently altered, \texttt{UpdateDecaySummary} must be called to
ensure consistency of the summary list.\\

\subsection{The \texttt{TTMParticleSet} Class}

The thermalised fireballs considered in statistical-thermal models typically contain 
approximately 350 different hadron and hadronic resonance species. To facilitate fast 
retrieval of particle properties, the \texttt{TTMParticle} objects of all constituents 
are stored in a hash table 
in a \texttt{TTMParticleSet} object. Other data members of this \texttt{TTMParticleSet} class 
include the filename used to instantiate the object and the number of particle species. 
Access to the entries in the hash table is through the particle Monte Carlo ID's.\\ 

\subsubsection{Instantiating a \texttt{TTMParticleSet} Object}

In addition to the default constructor, the following constructors exist:

\begin{center}
\begin{tabular}{l}
\texttt{TTMParticleSet *set = new TTMParticleSet(char *file);}\\
\texttt{TTMParticleSet *set = new TTMParticleSet(TDatabasePDG *pdg);}
\end{tabular}
\end{center}
 
The first constructor instantiates a \texttt{TTMParticleSet} object and inputs the particle properties 
contained in the specified text file. As an example of such a file, 
\texttt{/\$THERMUS/particles/PartList\_PPB2002.txt} contains a list 
of all mesons (up to the $K_4^*(2045)$) and baryons (up to the $\Omega^-$) 
listed in the July 2002 Particle Physics Booklet~\cite{PDG} (195 entries). Only particles 
need be included, since the anti-particle properties are directly related to 
those of the corresponding particle. The required file format is as follows:

\small
\begin{verbatim}

0       Delta(1600)0    32114   4       +1      1.60000      0       1       
0      0     0.35000    1.07454 (npi0)

\end{verbatim}
\normalsize

\begin{itemize}
\item{stability flag (1 for stable, 0 for unstable)}
\item{particle name}
\item{Monte Carlo particle ID (used for all referencing)}
\item{spin degeneracy}
\item{statistics (+1 for Fermi-Dirac, -1 for 
Bose-Einstein, 0 for Boltzmann)}
\item{mass in GeV}
\item{strangeness}
\item{baryon number}
\item{charge}
\item{absolute strangeness content $\left|{S}\right|$}
\item{width in GeV}
\item{threshold in GeV}
\item{string recording the decay channel from which the threshold is calculated if the particle's width is non-zero}
\end{itemize}

All further particle properties have to be set with the relevant `setters' (e.g. the charm, absolute charm 
content and hard-sphere radius). By default, all properties not listed in the particle list 
file are assumed to be zero.\\
 
\begin{figure}
\begin{center}
\includegraphics[width=12cm]{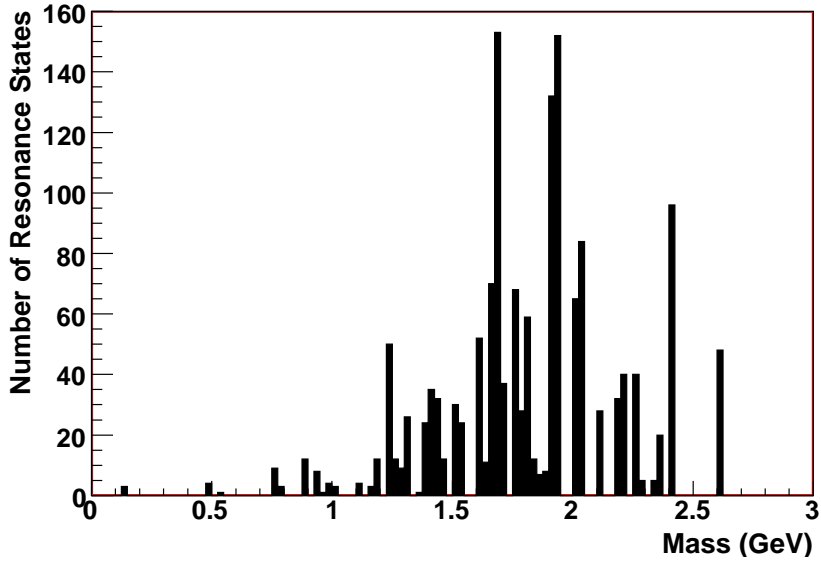}
\caption{The mass distribution 
of the resonances included in \texttt{PartList\_PPB2002.txt}.}\label{ResDistr}
\end{center}
\end{figure}

Figure \ref{ResDistr} shows the distribution of resonances (both particle and 
anti-particle) derived from \texttt{/\$THERMUS/particles/PartList\_PPB2002.txt}. 
As collider energies increase, so does the need to include also the higher mass 
resonances. Although the \texttt{TTMParticle} class allows for the 
properties of charmed particles, these particles are not included in the default THERMUS 
particle list. If required, these particles have to be input by the user. The same applies to 
the hadrons composed of $b$ and $t$ quarks.\\

It is also possible to use a \texttt{TDatabasePDG} object to instantiate a particle set\footnote{In order 
to have access to \texttt{TDatabasePDG} and related classes, one must first load 
\texttt{/\$ROOTSYS/lib/libEG.so}}. \texttt{TDatabasePDG} objects also read in particle information from text 
files. The default file is \texttt{/\$ROOTSYS/etc/pdg\_table.txt} and is based on the parameters used in PYTHIA \cite{Pythia}.\\

The constructor \texttt{TTMParticleSet(TDatabasePDG *pdg)} extracts only those particles in 
the specified \texttt{TDatabasePDG} object in particle classes `Meson', `CharmedMeson', `HiddenCharmMeson', 
`B-Meson', `Baryon', `CharmedBaryon' and `B-Baryon', as specified in \texttt{/\$ROOTSYS/etc/pdg\_table.txt}, 
and includes them in the hadron set. Anti-particles must be included in the \texttt{TDatabasePDG} object, 
as they are not automatically generated in this constructor of the \texttt{TTMParticleSet} class.\\

The default file read into the \texttt{TDatabasePDG} object, however, is incomplete; the charm, degeneracy, 
threshold, strangeness, $\left|S\right|$, beauty and topness of the particle are not included. 
Although the \texttt{TDatabasePDG::ReadPDGTable} function and default file allow for isospin, 
$I_3$, spin, flavor and tracking code to be entered too, the default file does not contain these values. 
Furthermore, all particles are made stable by default. Therefore, at present, using the 
\texttt{TDatabasePDG} 
class to instantiate a \texttt{TTMParticleSet} class should be avoided, at least until \texttt{pdg\_table.txt} is improved. 

\subsubsection{Inputting Decays}

Once a particle set has been defined, the decays to the stable particles 
in the set can be determined. After instantiating a \texttt{TTMParticleSet} 
object and settling on its stable constituents (the list of stable particles 
can be modified by adjusting the stability flags 
of the \texttt{TTMParticle} objects included in the \texttt{TTMParticleSet} object), 
decays can be input using the \texttt{InputDecays} method. 
Running this function populates the decay lists of 
all unstable particles in the set, using the decay files listed in the directory 
specified as the first argument. If a file is not found, then the corresponding particle 
is set to stable. For each typically unstable particle in 
\texttt{/\$THERMUS/particles/PartList\_PPB2002.txt}, there exists a file in 
\texttt{/\$THERMUS/particles} listing its decays. The 
filename is derived from the particle's name (e.g. \texttt{Delta(1600)0\_decay.txt} 
for the $\Delta(1600)^0$). There are presently 195 such files, with entries 
based on the Particle Physics Booklet of July 2002~\cite{PDG}. The decays of the 
corresponding anti-particles are 
automatically generated, while a private recursive function, \texttt{GenerateBRatios}, 
is invoked to ensure that only stable particles feature in the decay summary lists. The second argument of \texttt{InputDecays}, when set to true, scales the 
branching ratios so that their sum is 100\%. As an example, consider the 
following (again only part of the listing is shown):\\

\small
\begin{verbatim}

root [ ] TTMParticleSet set("$THERMUS/particles/PartList_PPB2002.txt")
root [ ] set.InputDecays("$THERMUS/particles/",true)
root [ ] TTMParticle *part = set.GetParticle(32114)
root [ ] part->List()

         ********* LISTING FOR PARTICLE Delta(1600)0 *********

                                     -
                                     -
                                     -

                 UNSTABLE

                 Decay Channels:
                 BRatio: 0.108558          Daughters:      2112    111
                 BRatio: 0.0542326         Daughters:      2212    -211
                 BRatio: 0.272837          Daughters:      2214    -211
                 BRatio: 0.0341395         Daughters:      2114    111
                 BRatio: 0.204651          Daughters:      1114    211
                 BRatio: 0.0774884         Daughters:      2112    113
                 BRatio: 0.0387907         Daughters:      2212    -213
                 BRatio: 0.139535          Daughters:      12112   111
                 BRatio: 0.0697674         Daughters:      12212   -211


                 Summary of Decays:
                2112            60.6774%
                111             62.5704%
                2212            39.3226%
                -211            83.9999%
                211             44.6773%
         **********************************************

\end{verbatim}
\normalsize

For particle sets based on \texttt{TDatabasePDG} objects, decay lists should be populated through the function 
\texttt{InputDecays(TDatabasePDG *)}. This function, however, does not automatically generate 
the anti-particle decays from those of the particle. Instead, the anti-particle decay list is used. 
Since the decay list may include electromagnetic and weak decays to particles other than the 
hadrons stored in the \texttt{TTMParticleSet} object, each channel is first checked to ensure that it 
contains only particles listed in the set. If not, the channel is excluded from the hadron's decay 
list used by THERMUS. As mentioned earlier, care should be taken when using \texttt{TDatabasePDG} objects 
based on the default file, as it is incomplete.\\ 

An extremely useful function is \texttt{ListParents(Int\_t id)}, which lists all 
of the parents of the particle with the specified Monte Carlo ID. This function uses 
\texttt{GetParents(TList *parents, Int\_t id)}, which populates the list passed with 
the decays to particle \texttt{id}. Note that these parents are 
not necessarily `direct parents'; the decays may involve unstable intermediates. 

\subsubsection{Customising the Set}

The \texttt{AddParticle} and \texttt{RemoveParticle} functions allow customisation of particle sets. 
Particle and anti-particle are treated symmetrically 
in the case of the former; if a particle is added, then its corresponding anti-particle is also 
added. This is not the case for the \texttt{RemoveParticle} function, however, where particle 
and anti-particle have to be removed separately.\\ 

Mass-cuts can be performed using 
\texttt{MassCut(Double\_t x)} to exclude 
all hadrons with masses greater than the argument (expressed in GeV). Decays then have to be re-inserted, 
to remove the influence of the newly-excluded hadrons from the decay lists.\\ 

The function \texttt{SetDecayEfficiency} allows the reconstruction efficiency 
of the decays from a specified parent to the specified daughter to be set. 
Changes are reflected only in the decay summary list of the parent (i.e. not 
the decay channel list). Note that running 
\texttt{UpdateDecaySummary} or \texttt{GenerateBRatios} will remove any such 
changes, by creating again a summary list consistent with the channel list.\\    

In addition to these operations, users can input their own particle sets by 
compiling their own particle lists and decay files.\\

\subsection{The \texttt{TTMParameter} Class}

This class groups all relevant information for parameters in the statistical-thermal model. Data members include:\\

\begin{tabular}{ll}
\texttt{fName} &- the parameter name,\\
\texttt{fValue} &- the parameter value,\\
\texttt{fError} &- the parameter error,\\
\texttt{fFlag} &- a flag signalling the type of parameter (constrain, fit,\\
               &  fixed, or uninitialised),\\
\texttt{fStatus} &- a string reflecting the intended treatment or action taken.
\end{tabular}\\

\noindent
In addition to these data members, the following, relevant to fit-type parameters, are also included:\\

\begin{tabular}{ll}
\texttt{fStart} &- the starting value in a fit,\\
\texttt{fMin} &- the lower bound of the fit-range,\\
\texttt{fMax} &- the upper bound of the fit-range,\\
\texttt{fStep} &- the step-size.
\end{tabular}\\

\noindent
The constructor and \texttt{SetParameter(TString name, Double\_t value, Double\_t error)} function set the parameter 
to fixed-type, by default. The parameter-type can be modified using the \texttt{Constrain}, 
\texttt{Fit} or \texttt{Fix} methods.

\subsection{The \texttt{TTMParameterSet} Class}
 
The \texttt{TTMParameterSet} class is the base class for all thermal parameter set 
classes. Each derived class contains its own \texttt{TTMParameter} array, with size determined 
by the requirements of the ensemble. The base class contains a pointer to 
the first element of this array. In addition, it stores the constraint 
information.\\

All derived classes contain the function 
\texttt{GetRadius}. In this way, \texttt{TTMParameterSet} is able to define a function, \texttt{GetVolume}, 
which returns the volume required to convert densities into total fireball quantities.\\
 
\texttt{TTMParameterSetBSQ}, \texttt{TTMParameterSetBQ} and \texttt{TTMParameterSetCanBSQ} are the derived classes.\\ 

\subsubsection{\texttt{TTMParameterSetBSQ}}
This derived class, applicable to the grand-canonical ensemble, contains the parameters:\\

\begin{center}
\begin{tabular}{cccccccc}
$T$ & $\mu_B$ & $\mu_S$ & $\mu_Q$ & $\mu_C$ & $\gamma_S$ & $\gamma_C$ & $R$,\\
\end{tabular}
\end{center}

\noindent
where $R$ is the fireball radius, assuming a spherical fireball (i.e. $V=4/3\pi R^3$). 
In addition, the $B/2Q$ ratio and charm and strangeness density of the 
system are stored here. In the constructor, all errors are defaulted to zero, as is $R$, $\mu_C$, $S/V$, 
$C/V$ and $B/2Q$, while $\gamma_C$ is defaulted to unity.\\ 

Each parameter has a `getter' (e.g. \texttt{GetTPar}), which returns a pointer to the requested  
\texttt{TTMParameter} object. In this class, 
$\mu_S$ and $\mu_Q$ can be set to constrain-type using \texttt{ConstrainMuS} and 
\texttt{ConstrainMuQ}, where the arguments are the required strangeness density and 
$B/2Q$ ratio, respectively. No such function exists for $\mu_C$, since constraining functions 
are not yet coded for the charm density. Each parameter of this class can be set to 
fit-type, using functions such as \texttt{FitT} (where the fit parameters have reasonable 
default values), or fixed-type, using functions such as \texttt{FixMuB}.
  
\subsubsection{\texttt{TTMParameterSetBQ}}
This derived class, applicable to the strangeness-canonical ensemble (strangeness exactly conserved and 
$B$ and $Q$ treated grand-canonically), has the parameters:\\
\begin{center}
\begin{tabular}{cccccc}
$T$ & $\mu_B$ & $\mu_Q$ & $\gamma_S$ & $R_c$ & $R$,\\
\end{tabular}
\end{center}

\noindent
where $R_c$ is the canonical or correlation radius; the radius inside 
which strangeness is exactly conserved. The fireball radius $R$, on the other hand, is used to 
convert densities into total 
fireball quantities. In addition, the required $B/2Q$ 
ratio is also stored, as well as the strangeness required inside the
correlation volume (which must be an integer).\\ 

In addition to the same `getters' and `setters' as the previous derived class, it is possible to set 
$\mu_Q$ to constrain-type by specifying the $B/2Q$ ratio in the argument of 
\texttt{ConstrainMuQ}. The strangeness required inside the canonical
volume is set through the \texttt{SetS} method. This value is
defaulted to zero. The function \texttt{ConserveSGlobally} fixes the 
canonical radius, $R_c$, to the 
fireball radius, $R$. As in the case of the \texttt{TTMParameterSetBSQ} class, there also exist 
functions to set each parameter to fit or fixed-type.\\

\subsubsection{\texttt{TTMParameterSetCanBSQ}}
This set, applicable to the canonical ensemble with exact conservation of $B$, $S$ and $Q$, contains 
the parameters:\\
\begin{center}
\begin{tabular}{cccccc}
$T$ & $B$ & $S$ & $Q$ & $\gamma_S$ & $R$.\\
\end{tabular}
\end{center}

\noindent
Since all conservation is exact, there are no chemical potentials to satisfy constraints. Again, the same 
`getters', `setters' and functions to set each parameter to fit or fixed-type exist, as in the case 
of the previously discussed \texttt{TTMParameterSet} derived classes.\\

\subsubsection{Example}

As an example, let us define a \texttt{TTMParameterSetBQ} object. By default, all 
parameters are initially of fixed-type. Suppose we wish to fit $T$ and 
$\mu_B$, and use $\mu_Q$ to constrain the $B/2Q$ ratio in the model to that 
in Pb+Pb collisions:

\small
\begin{verbatim}

root [ ] TTMParameterSetBQ parBQ(0.160,0.2,-0.01,0.8,6.,6.)
root [ ] parBQ.FitT(0.160)
root [ ] parBQ.FitMuB(0.2)
root [ ] parBQ.ConstrainMuQ(1.2683)
root [ ] parBQ.List()
  ***************************** Thermal Parameters ****************************

                 Strangeness inside Canonical Volume = 0

       T         =          0.16                    (to be FITTED)
                                                     start: 0.16
                                                     range: 0.05 -- 0.18
                                                     step:  0.001

     muB         =           0.2                    (to be FITTED)
                                                     start: 0.2
                                                     range: 0 -- 0.5
                                                     step:  0.001

     muQ         =         -0.01                    (to be CONSTRAINED)

                                                     B/2Q: 1.2683

  gammas         =           0.8                    (FIXED)

Can. radius      =             6                    (FIXED)

  radius         =             6                    (FIXED)

                         Parameters unconstrained

  ******************************************************************************

\end{verbatim}
\normalsize

\noindent
Note the default parameters for the $T$ and $\mu_B$ fits. Obviously, no constraining or fitting can 
take place yet; we have simply signalled our intent to take these actions at some later stage.

\subsection{The \texttt{TTMThermalParticle} Class}

By combining a \texttt{TTMParticle} and \texttt{TTMParameterSet} object, a thermal particle can be 
created. The 
\texttt{TTMThermalParticle} class is the base class from which thermal particle classes relevant 
to the three currently implemented thermal model formalisms, \texttt{TTMThermalParticleBSQ}, 
\texttt{TTMThermalParticleBQ} and \linebreak
\texttt{TTMThermalParticleCanBSQ}, are derived. Since no particle set is 
specified, the total fireball properties cannot be determined. Thus, in the grand-canonical 
approach, the constraints cannot yet be imposed to determine the values of the chemical potentials of constrain-type, while, in the strangeness-canonical and 
canonical formalisms, the canonical correction factors cannot yet be calculated. Instead, at this 
stage, the chemical potentials and/or correction factors must be specified.\\ 

Use is made of the fact that, in the Boltzmann approximation, $E/V$, $N/V$ and $P$, in the canonical 
and strangeness-canonical ensembles, are simply the grand-canonical values, with the chemical 
potential(s) corresponding to the canonically-treated quantum number(s) set to zero, multiplied by a 
particle-specific correction factor. This allows the functions for calculating $E/V$, $N/V$ and $P$ in 
the Boltzmann approximation to be included in the base class, which then also contains the correction 
factor as a data member (by definition, this correction factor is 1 in the grand-canonical ensemble).\\ 

Both functions including and excluding resonance width, $\Gamma$, are coded 
(e.g. \texttt{DensityBoltzmannNoWidth} and \texttt{EnergyBoltzmannWidth}). 
When width is included, a Breit-Wigner 
distribution is integrated over between the limits \linebreak 
$[\mathrm{max}(m-2\Gamma,m_{\mathrm{threshold}}),m+2\Gamma]$.\\

\subsubsection{\texttt{TTMThermalParticleBSQ}}

This class is relevant to the grand-canonical treatment of $B$, $S$ and $Q$. In
 addition to the functions for calculating $E/V$, $N/V$ and $P$ in the Boltzmann approximation, defined 
in the base class, functions implementing quantum statistics for these quantities exist in this 
derived class (e.g. \texttt{EnergyQStatNoWidth} and \texttt{PressureQStatWidth}). Additional member 
functions of this class calculate the entropy using either
 Boltzmann or quantum statistics, with or without width.\\ 

In the functions calculating the thermal quantities assuming quantum statistics, it is 
first checked that the integrals converge for the bosons (i.e. there is no Bose-Einstein 
condensation). The check is performed by the \texttt{ParametersAllowed} method. A warning 
is issued if there are problems and zero is returned.\\ 

This class also accommodates charm, since the associated parameter set includes 
$\mu_C$ and $\gamma_C$, while the associated particle may have non-zero charm.\\

\subsubsection{\texttt{TTMThermalParticleBQ}}

This class is relevant to the strangeness-canonical ensemble. At present, this 
class is only applied in the Boltzmann approximation. Under this assumption, 
$N/V$, 
$E/V$ and $P$ are given by the grand-canonical result, with $\mu_S$ set to zero, up to a 
multiplicative correction factor. Since the total 
entropy does not split into the sum of particle entropies, no entropy calculation is made in this 
class.\\
 
\subsubsection{\texttt{TTMThermalParticleCanBSQ}}

This class is relevant to the fully canonical treatment of $B$, $S$ and $Q$. At
 present, as in the case of \texttt{TTMThermalParticleBQ}, this class is only applied in the
 Boltzmann approximation. Also, since the total 
entropy again does not split into the sum of particle entropies, no entropy calculation is made here.\\

\subsubsection{Example}

Let us construct a thermal particle, within the strangeness-canonical ensemble, from the 
$\Delta(1600)^0$ and the parameter set previously defined. Since this particle has zero strangeness, 
a correction factor of 1 is passed as the third argument of the constructor:

\small
\begin{verbatim}
root [ ] TTMThermalParticleBQ therm_delta(part,&parBQ,1.)
root [ ] therm_delta.DensityBoltzmannNoWidth()
(Double_t)8.15072671710089913e-04
root [ ] therm_delta.EnergyBoltzmannWidth()
(Double_t)2.29185316377137748e-03
\end{verbatim}
\normalsize

\subsection{The \texttt{TTMThermalModel} Class}

Once a parameter and particle set have been specified, these can be 
combined into a thermal model. \texttt{TTMThermalModel} is the base class from which the 
\texttt{TTMThermalModelBSQ}, \texttt{TTMThermalModelBQ} and
\texttt{TTMThermalModelCanBSQ}\linebreak
classes are derived. A string descriptor is included as a data member of the base class 
to identify the type of model. This is used, for example, 
to handle the fact that the number of parameters in the associated parameter 
sets is different, depending on the model type.\\ 

All derived classes define functions 
to calculate the primordial particle, energy and entropy 
densities, as well as the pressure. These thermal quantities are stored in a hash table of \texttt{TTMDensObj} 
objects. Again, access is through the particle ID's. In addition to the individual particles' 
thermal quantities, the total primordial fireball strangeness, baryon, charge, charm, energy, entropy, 
and particle densities, pressure, and Wr\`oblewski factor (see Section~\ref{Wrob}) are included as data members.\\ 

At this level, the constraints on any chemical potentials of constrain-type can be imposed, and the 
correction factors in canonical treatments can be determined. Also, as soon as the 
primordial particle densities are known, the decay contributions can be calculated.\\ 

\subsubsection{Calculating Particle Densities}

Running \texttt{GenerateParticleDens} clears the current entries in the density hash table of the 
\texttt{TTMThermalModel} object, automatically constrains the chemical potentials (where applicable), 
calculates the canonical correction factors (where applicable), and then populates the density 
hash table with a \texttt{TTMDensObj} object for each particle in the associated set. 
The decay contributions to 
each stable particle are also calculated, so that the density hash table contains both primordial and 
decay particle density contributions, provided of course that the decays have been entered in the associated 
\texttt{TTMParticleSet} object. In addition, the 
Wr\`oblewski factor and total strangeness, baryon, charge, charm and particle densities in the fireball 
are calculated.\\ 

Note: The summary decay lists of the associated
\texttt{TTMParticleSet} object are used to calculate the decay contributions. Hence, only 
stable particles have decay contributions reflected in 
the hash table. Unstable particles that are themselves fed by higher-lying 
resonances, do not receive a decay contribution.\\

Each derived class contains the private function \texttt{PrimPartDens}, which calculates only the primordial 
particle densities and, hence, the canonical correction factors, where applicable. In the case of 
the grand-canonical and strangeness-canonical ensembles, this function calculates the densities without 
automatically constraining the chemical potentials of constrain-type first. The constraining is handled by 
\texttt{GenerateParticleDens}, which calls external friend functions, which, in turn, call 
\texttt{PrimPartDens}. In the purely canonical ensemble, \texttt{GenerateParticleDens} simply calls 
\texttt{PrimPartDens}. In this way, there is uniformity between the derived classes. Since there is no 
constraining to be done, there is no real need for a separate function in the canonical case.\\  

\subsubsection{Calculating Energy and Entropy Densities and Pressure}

\texttt{GenerateEnergyDens}, \texttt{GenerateEntropyDens} and \texttt{GeneratePressure} 
iterate through the existing density 
hash table and calculate and insert, respectively, the primordial energy density, entropy density and pressure of each 
particle in the set. In addition, they calculate the total primordial energy density, entropy 
density and pressure in the fireball, respectively. These functions require that the density hash table 
already be in existence. In other 
words, \texttt{GenerateParticleDens} must already have been run. If the parameters have subsequently changed, 
then this function must be run yet again to recalculate the correction factors or re-constrain the parameters, 
as required.\\

\subsubsection{Bose-Einstein Condensation}

When quantum statistics are taken into account (e.g. in \texttt{TTMThermalModelBSQ} or 
for the non-strange particles in \texttt{TTMThermalModelBQ}), certain choices of 
parameters lead to diverging 
integrals for the bosons (Bose-Einstein condensation). 
In these classes, a check, based on 
\texttt{TTMThermalParticleBSQ::Parameters-} \texttt{Allowed}, is included to ensure that the parameters 
do not lead to problems. Including also the possibility of incomplete strangeness and/or charm 
saturation (i.e. $\gamma_S\neq1$ and/or $\gamma_C\neq1$), Bose-Einstein condensation is avoided, 
provided that,
\begin{equation}
e^{\left(m_i-\mu_i\right)/T} > \gamma_S^{\left|S_i\right|}\gamma_C^{\left|C_i\right|},
\end{equation}
for each boson. If this condition is failed to be met for any of the bosons in the set, a 
warning is issued and the densities are not calculated.\\

\subsubsection{Accessing the Thermal Densities}

The entries in the density hash table are accessed using the particle Monte Carlo ID's. The function 
\texttt{GetDensities(Int\_t ID)} returns the \texttt{TTMDensObj} object containing the thermal quantities 
of the particle 
with the specified ID. The primordial particle, energy, and entropy densities, pressure, and decay density 
are extracted from this object using the \texttt{GetPrimDensity}, \texttt{GetPrimEnergy}, 
\texttt{GetPrimEntropy}, \texttt{GetPrimPressure}, and \texttt{GetDecayDensity} functions of the \texttt{TTMDensObj} class, respectively. 
The sum of the primordial and decay particle 
densities is returned by \texttt{TTMDensObj::GetFinalDensity}. \texttt{TTMDensObj::List} outputs to 
screen all thermal densities 
stored in a\linebreak
\texttt{TTMDensObj} object.\\
 
\texttt{ListStableDensities} lists the densities (primordial and decay contributions) of all those particles 
considered stable in the particle set associated with the model. Access to the total fireball densities is through separate `getters' defined in the \texttt{TTMThermalModel} 
base class (e.g. \texttt{GetStrange}, \texttt{GetBaryon} etc.).\\ 

\subsubsection{Further Functions}

\texttt{GenerateDecayPartDens} and \texttt{GenerateDecayPartDens(Int\_t id)} 
 (both defined in the 
base class) calculate 
decay contributions to stable particles. The former iterates through the density hash table and calculates 
the decay contributions to all those particles considered stable in the set. The latter calculates 
just the contribution to the stable particle with the specified ID. In both cases, the primordial densities 
must be 
calculated first. In fact, \texttt{GenerateParticleDens} automatically calls 
\texttt{GenerateDecayPartDens}, so that this 
function does not have to be run separately under ordinary circumstances. However, if one is interested 
in investigating the effect of decays, while keeping the parameters (and hence the primordial densities) 
fixed, then running these functions is best (the hash table will not be repeatedly cleared and repopulated 
with the same primordial densities).\\

\texttt{ListDecayContributions(Int\_t d\_id)} lists the contributions (in percentage and absolute terms) of 
decays to the daughter with the specified ID. The primordial and decay densities must already appear in 
the density 
hash table (i.e. run \texttt{GenerateParticleDens} first). \texttt{ListDecayContribution(Int\_t p\_id,
Int\_t d\_id)} lists the contribution of the decay from the specified parent (with ID \texttt{p\_id}) to 
the specified daughter (with ID \texttt{d\_id}). The percentages listed by each of these functions are 
those of the individual decays to the total decay density.\\

Next we consider the specific features of the derived \texttt{TTMThermalModel} 
classes.

\subsubsection{\texttt{TTMThermalModelBSQ}}

In the grand-canonical ensemble, quantum statistics can be employed and, hence, there is a flag specifying 
whether to use Fermi-Dirac and Bose-Einstein statistics or Boltzmann statistics. The constructor, by default, includes both the effect of quantum statistics and resonance width. 
The flags controlling their inclusion are set using the 
\texttt{SetQStats} and \texttt{SetWidth} functions, respectively. The 
functions that calculate the particle, energy, and entropy densities, and pressure then use the 
corresponding functions in the \texttt{TTMThermalParticleBSQ} class to calculate these quantities in the 
required way. The statistics data member (\texttt{fStat}) of each 
\texttt{TTMParticle} included in the associated set can be used to fine-tune 
the inclusion of quantum statistics; with the quantum statistics flag switched 
on, Boltzmann statistics are still used for those particles with 
\texttt{fStat=0}.\\ 

In this ensemble, at this stage, both $\mu_S$ and $\mu_Q$ can be constrained (either separately or 
simultaneously). In order to accomplish this, the $\mu_S$ and/or $\mu_Q$ parameters in the associated 
\texttt{TTMParameterSetBSQ} object must be set to constrain-type.\\

It is also possible to constrain $\mu_B$ by the primordial ratio $E/N$ (the average energy per 
hadron), $n_b+n_{\bar{b}}$ (the total primordial baryon plus anti-baryon density), or $s/T^3$ (the 
primordial, temperature-normalised entropy density). 
This is accomplished by the $\texttt{ConstrainEoverN}$, $\texttt{ConstrainTotalBaryonDensity}$ and 
$\texttt{ConstrainSoverT3}$ methods, respectively. Running these functions will adjust $\mu_B$ such 
that $E/N$, $n_b+n_{\bar{b}}$ or $s/T^3$, respectively, has the required value, regardless of the 
parameter type of $\mu_B$. In addition, the percolation model \cite{perc} can be imposed 
to constrain $\mu_B$ using \texttt{ConstrainPercolation}.\\
  
This class also accommodates charm, since the associated parameter set includes 
$\mu_C$ and $\gamma_C$, while the associated particle set may contain charmed particles. However, 
no constraining functions have yet been written for the charm content within this ensemble.\\  

Within the grand-canonical ensemble, it is possible to include excluded volume effects. Their 
inclusion is controlled by 
the \texttt{fExclVolCorrection} flag, false by default, which is set through the 
\texttt{SetExcludedVolume} function. When included, these corrections are calculated on calling 
{\tt GenerateParticleDens}, based on the hard-sphere radii stored in the \texttt{TTMParticle} objects 
of the associated particle set.\\ 

\subsubsection{\texttt{TTMThermalModelBQ}}
This class contains the following additional data members:\\

\begin{tabular}{ll}
\texttt{flnZtot}   &- log of the total partition function,\\
\texttt{flnZ0}     &- log of the non-strange component of the partition function,\\
\texttt{fExactMuS} &- equivalent strangeness chemical potential,\\
\texttt{fCorrP1}   &- canonical correction for S = +1 particles,\\
\texttt{fCorrP2}   &- canonical correction for S = +2 particles,\\
\texttt{fCorrP3}   &- canonical correction for S = +3 particles,\\
\texttt{fCorrM1}   &- canonical correction for S = -1 particles,\\
\texttt{fCorrM2}   &- canonical correction for S = -2 particles,\\
\texttt{fCorrM3}   &- canonical correction for S = -3 particles.
\end{tabular}\\

Although this ensemble is only applied in the Boltzmann approximation for $S\neq0$ hadrons, it is 
possible to apply quantum statistics to the $S=0$ hadrons. This is achieved through the 
\texttt{SetNonStrangeQStats} function. By default, quantum statistics is included for the 
non-strange hadrons by the constructors. Resonance width can be included for all hadrons, 
and is achieved through the \texttt{SetWidth} function. The constructors, by default, apply 
resonance width. The functions that calculate the particle, energy, and entropy densities, and 
pressure then use the corresponding functions in the \texttt{TTMThermalParticle} classes to calculate 
these quantities in the required way.\\ 

\texttt{GenerateParticleDens} populates the density hash table with particle densities, including 
the canonical correction factors, which are also stored in the appropriate data members. The 
equivalent strangeness chemical potential is calculated from the canonical
correction factor for $S=+1$ particles. In the limit of large $VT^3$, 
this approaches the value of $\mu_S$ in the equivalent grand-canonical treatment.\\

Running \texttt{GenerateEntropyDens} populates each \texttt{TTMDensObj} object in the hash table 
with only that part of the total entropy that can be unambiguously attributed to that particular 
particle. There is a term in the total entropy that cannot be split; this 
is added to the total entropy at the end, but not included in the individual entropies (i.e. summing up 
the entropy contributions of each particle will not give the total entropy).\\

At this stage, in this formalism, $\mu_Q$ can be constrained 
(this is automatically realised if this parameter is set to constrain-type), while the 
correlation radius ($R_c$) can be set to the fireball radius ($R$) by applying the function 
\texttt{ConserveSGlobally} to the associated \texttt{TTMParameterSetBQ} object.\\

In exactly the same way as in the grand-canonical ensemble case, $\mu_B$ can 
be constrained in this ensemble by the primordial ratio $E/N$ (the average 
energy per 
hadron), $n_b+n_{\bar{b}}$ (the total primordial baryon plus anti-baryon 
density), or $s/T^3$ (the primordial, temperature-normalised entropy density), as well 
as by the percolation model.\\
 
\subsubsection{\texttt{TTMThermalModelCanBSQ}}

This class contains, amongst others, the following data members:\\

\begin{tabular}{ll}
\texttt{flnZtot}         &- log of the total canonical partition function,\\  
\texttt{fMuB,fMuS,fMuQ}  &- equivalent chemical potentials,\\
\texttt{fCorrpip}        &- correction for $\pi^+$-like particles,\\
\texttt{fCorrpim}        &- correction for $\pi^-$-like particles,\\
\texttt{fCorrkm}         &- correction for $K^-$-like particles,\\
\texttt{fCorrkp}         &- correction for $K^+$-like particles,\\
\texttt{fCorrk0}         &- correction for $K^0$-like particles,\\
\texttt{fCorrak0}        &- correction for $\bar{K}^0$-like particles,\\
\texttt{fCorrproton}     &- correction for $p$-like particles,\\
\texttt{fCorraproton}    &- correction for $\bar{p}$-like particles,\\
\texttt{fCorrneutron}    &- correction for $n$-like particles,\\
\texttt{fCorraneutron}   &- correction for $\bar{n}$-like particles,\\
\texttt{fCorrlambda}     &- correction for $\Lambda$-like particles,\\
\texttt{fCorralambda}    &- correction for $\bar{\Lambda}$-like particles,\\
\texttt{fCorrsigmap}     &- correction for $\Sigma^+$-like particles,\\
\texttt{fCorrasigmap}    &- correction for $\bar{\Sigma}^-$-like particles,\\
\texttt{fCorrsigmam}     &- correction for $\Sigma^-$-like particles,\\
\texttt{fCorrasigmam}    &- correction for $\bar{\Sigma}^+$-like particles,\\
\texttt{fCorrdeltam}     &- correction for $\Delta^-$-like particles,
\end{tabular}

\begin{tabular}{ll}
\texttt{fCorradeltam}    &- correction for $\bar{\Delta}^+$-like particles,\\
\texttt{fCorrdeltapp}    &- correction for $\Delta^{++}$-like particles,\\
\texttt{fCorradeltapp}   &- correction for $\bar{\Delta}^{--}$-like particles,\\
\texttt{fCorrksim}       &- correction for $\Xi^{-}$-like particles,\\
\texttt{fCorraksim}      &- correction for $\bar{\Xi}^+$-like particles,\\
\texttt{fCorrksi0}       &- correction for $\Xi^0$-like particles,\\
\texttt{fCorraksi0}      &- correction for $\bar{\Xi}^0$-like particles,\\
\texttt{fCorromega}      &- correction for $\Omega^-$-like particles,\\
\texttt{fCorraomega}     &- correction for $\bar{\Omega}^+$-like particles.
\end{tabular}\\

Since this ensemble is only applied in the Boltzmann approximation, there is no flag for quantum 
statistics. However, resonance width can be included. This is achieved through the 
\texttt{SetWidth} function. The constructor, by default, applies resonance width. The 
functions that calculate the particle, energy, and entropy densities, and pressure then use the 
corresponding functions in the \texttt{TTMThermalParticle} classes to calculate these 
quantities in the required way.\\ 

\texttt{GenerateParticleDens} calls \texttt{PrimPartDens}, which calculates the particle densities, 
including the canonical correction factors, which are then also stored in the relevant data members 
accessible through the \texttt{GetCorrFactor} method. The integrands featuring in the evaluation 
of the partition function and correction factors can be viewed after calling 
\texttt{PopulateZHistograms}. This function populates the array passed as argument with 
histograms showing these integrands as a function of the integration variables $\phi_S$ and 
$\phi_Q$. Since these histograms 
are created off of the heap, they must be cleaned up afterwards.\\ 

\texttt{GenerateEntropyDens} acts in exactly the same way as in the strangeness-canonical ensemble case.\\

\subsubsection{Example}

As an example, we consider the strangeness-canonical ensemble, based on the particle set and 
strangeness-canonical parameter set previously defined. After instantiating the object, we populate the hash table 
with primordial and decay particle densities:

\small
\begin{verbatim}

root [ ] TTMThermalModelBQ modBQ(&set,&parBQ)
root [ ] modBQ.GenerateParticleDens()
root [ ] parBQ.List()

  ***************************** Thermal Parameters ****************************

                 Strangeness inside Canonical Volume = 0

       T         =          0.16                    (to be FITTED)
                                                     start: 0.16
                                                     range: 0.05 -- 0.18
                                                     step:  0.001

     muB         =           0.2                    (to be FITTED)
                                                     start: 0.2
                                                     range: 0 -- 0.5
                                                     step:  0.001

     muQ         =       -0.00636409                (*CONSTRAINED*)

                                                     B/2Q: 1.2683

  gammas         =           0.8                    (FIXED)

Can. radius      =             6                    (FIXED)

  radius         =             6                    (FIXED)

                         B/2Q Successfully Constrained

  ******************************************************************************
\end{verbatim}
\normalsize

\noindent 
One notices that the constraint on $\mu_Q$ is now automatically 
imposed.\\ 

The energy and entropy densities and pressure can be calculated once \texttt{GenerateParticleDens} 
has been run:

\small
\begin{verbatim}
root [ ] modBQ.GenerateEnergyDens()
root [ ] modBQ.GenerateEntropyDens()
root [ ] modBQ.GeneratePressure()
\end{verbatim}
\normalsize

\noindent
Now, suppose that we are interested in the thermal densities of the 
$\Delta(1600)^0$ and $\pi^+$:

\small
\begin{verbatim}

root [ ] TTMDensObj *delta_dens = modBQ.GetDensities(32114)
root [ ] delta_dens->List()
  **** Densities for Particle 32114 ****
         n_prim = 0.00138306
         n_decay = 0
         e_prim = 0.0022912
         s_prim = 0.0139745
         p_prim = 0.000221328

root [ ] TTMDensObj *piplus_dens = modBQ.GetDensities(211)
root [ ] piplus_dens->List()
  **** Densities for Particle 211 ****
         n_prim = 0.0488139
         n_decay = 0.119683
         e_prim = 0.0247039
         s_prim = 0.20276
         p_prim = 0.00742708

\end{verbatim}
\normalsize

\noindent
One notices that the $\pi^+$ has a decay density contribution, while the $\Delta(1600)^0$ does not. 
This is because, unlike the $\Delta(1600)^0$, the $\pi^+$ was considered stable.\\

\subsubsection{Imposing of Constraints}

The `Numerical Recipes in C'~\cite{NRC} function applying the Broyden globally
convergent secant method of solving nonlinear systems of equations is
employed by THERMUS to constrain parameters. The input to the Broyden 
method is a vector of functions for which roots are sought. Typically, 
in the thermal model, solutions to the following equations 
are required (either separately or simultaneously):\\  
\begin{eqnarray*}
\left({B/V \over 2Q/V}\right)_{primordial}^{model} - \left({B \over 2Q}\right)^{colliding\;system} &=& 0,\\
S_{primordial}^{model} - S^{colliding\;system} &=& 0,\\
\left({E/V \over N/V}\right)_{primordial}^{model} - \left(E \over N\right)^{required} &=& 0.
\end{eqnarray*}

Although, as written, these equations are correct, the quantities $B/2Q$, $S$ and 
$E/N$ are typically of different orders of magnitude. Since the Broyden method in `Numerical 
Recipes in C' defines just one tolerance level for function convergence (\texttt{TOLF}), it 
is important to `normalise' each equation:

\begin{eqnarray*}
\left\{\left({B/V \over 2Q/V}\right)_{primordial}^{model} - \left({B \over 2Q}\right)^{colliding\;system}\right\} / \left({B \over 2Q}\right)^{colliding\;system}  &=& 0,\\
\left\{S_{primordial}^{model} - S^{colliding\;system}\right\} / S^{colliding\;system} &=& 0,\\
\left\{\left({E/V \over N/V}\right)_{primordial}^{model} - \left({E \over N}\right)^{required}\right\} / \left({E \over N}\right)^{required}&=& 0.
\end{eqnarray*}

This is the most democratic way of treating the constraints. However, this method obviously fails 
in the event of one of the denominators being zero. For the equations considered above, this is only 
likely in the case of the strangeness constraint, where the initial strangeness content is typically zero. In 
this 
case, where the strangeness carried by the positively strange particles $S_+$ is balanced by the 
strangeness carried by the negatively strange particles $S_-$, we write as our function to be 
satisfied,

\begin{eqnarray*}
\left(S/V\right)_{primordial}^{model}/ \left(|S_+|_{primordial}^{model}/V+|S_-|_{primordial}^{model}/V\right) &=& 0.
\end{eqnarray*}

In this way, the constraints can be satisfied to equal relative degrees, and equally well fractionally 
at each point in the parameter space. In addition to the constraints listed above, THERMUS also allows for 
the constraining of the total baryon plus anti-baryon density and the temperature-normalised entropy 
density, $s/T^3$, as well as the imposing of the percolation model.\\

\subsubsection{Calculation of the Wr\`oblewski Factor}\label{Wrob}

The Wr\`oblewski factor~\cite{Wrob} is defined as,
\begin{eqnarray*}
\lambda_S &=& \frac{2<s\bar{s}>}{<u\bar{u}>+<d\bar{d}>},
\end{eqnarray*}
where $<u\bar{u}>+<d\bar{d}>$ is the sum of newly-produced $u\bar{u}$
and $d\bar{d}$ pairs, while all $s\bar{s}$ pairs are newly-produced if $S=0$ in the initial state.\\ 

In THERMUS, $\lambda_S$ is calculated in the following way:
\begin{itemize}

\item{Using the primordial particle densities and the strangeness content of each particle listed 
in the particle hash table, the $s + 
    \bar{s}$ and $u + d + \bar{u} + \bar{d}$ densities are determined.}
    \item{Assuming $S=0$, $\#s = \#\bar{s}$, and so the density of
    newly-produced $s\bar{s}$ pairs is simply $(s + \bar{s})/2$}.
  \item{From baryon number conservation, the net baryon content in the system, $n_B$, originates from the initial state. Thus, $3\times n_B$ must correspond to the density of
  $u+d$ quarks brought in by the colliding nuclei. This is subtracted from the total
  $u + d + \bar{u} + \bar{d}$ density to yield the density of newly-produced non-strange light quarks.}
\item{Since $\#s=\#\bar{s}$ and, amongst newly-produced non-strange light
  quarks, $u+d=\bar{u}+\bar{d}$, further assuming that $\mu_Q=0$
  implies that $u=\bar{u}=d=\bar{d}$. This allows the density of
  $u\bar{u}$ and $d\bar{d}$ pairs to be easily determined.}
\end{itemize}

\subsection{Thermal Fit Preliminaries}

Often a single experiment releases yields and ratios that contain different feed-down 
corrections. Each yield or ratio then has a different decay chain associated with it. Since \texttt{TTMThermalModel} objects allow for just one associated particle set, they do not allow sufficient flexibility for performing thermal fits to experimental 
data. However, \texttt{TTMThermalFit} classes do feature such flexibility. Before we discuss these classes, let us look at the \texttt{TTMYield} object, which forms an essential part of the \texttt{TTMThermalFit} class.\\

\subsubsection{\texttt{TTMYield}}

Information relating to both yields and ratios of yields can be stored in \texttt{TTMYield} objects. These 
objects contain the following data members:\\

\begin{tabular}{ll}
\texttt{fName} &- the name of the yield or ratio,\\
\texttt{fID1} &- the ID of the yield or numerator ID in the case of a ratio,\\
\texttt{fID2} &- denominator ID in the case of a ratio (0 for a yield),\\
\texttt{fFit} &- true if the yield or ratio is to be included in a fit (else predicted),\\
\texttt{fSet1} &- particle set relevant to yield or numerator in case of ratio,\\
\texttt{fSet2} &- particle set relevant to denominator in case of ratio (0 for yield),\\
\texttt{fExpValue} &- the experimental value,\\
\texttt{fExpError} &- the experimental error,\\
\texttt{fModelValue} &- the model value,\\
\texttt{fModelError} &- the model error.\\
\end{tabular}\\

By default, \texttt{TTMYield} objects are set for inclusion in fits. The 
functions \texttt{Fit} and \texttt{Predict} 
control the fit-status of a \texttt{TTMYield} object. Particle sets (decay chains) are assigned using the \texttt{SetPartSet} method.\\ 

The functions \texttt{GetStdDev} and \texttt{GetQuadDev} return the number 
of standard and quadratic deviations between model and experimental values, 
respectively, i.e., 
\begin{eqnarray}
(\mathrm{Model\;Value- Exp.\;Value})/\mathrm{Exp.\;Error},
\end{eqnarray}
and, 
\begin{eqnarray}
(\mathrm{Model\;Value - Exp.\;Value})/\mathrm{Model\;Value},
\end{eqnarray}
respectively, while \texttt{List} outputs the contents of a \texttt{TTMYield} 
object to screen. Access to all remaining data members is through the relevant `getters' and `setters'. \\ 

\subsection{The \texttt{TTMThermalFit} Class}

This is the base class from which the \texttt{TTMThermalFitBSQ}, \texttt{TTMThermalFitBQ} and \texttt{TTMThermalFitCanBSQ} classes are derived. Each \texttt{TTMThermalFit} object contains:
\begin{itemize}
\item{a particle set, the so-called base set, which contains all of the constituents of the hadron gas, as 
well as the default decay chain to be used;} 
\item{a parameter set;} 
\item{a list of \texttt{TTMYield} objects containing yields and/or ratios of interest;}
\item{data members storing the total $\chi^2$ and quadratic deviation; and} 
\item{a \texttt{TMinuit} fit object.}
\end{itemize}

A string descriptor is also included in the base class to 
identify the type of model on which the fit is based. This is used, for 
example, to determine the number of parameters in the associated 
parameter sets.\\ 

Each derived class defines a private function, \texttt{GenerateThermalModel}, which creates 
(off the heap) a thermal model object, based on the base particle set and parameter set of 
the \texttt{TTMThermalFit} object, with the specific quantum statistics/resonance width/excluded 
volume requirements, where applicable.\\

\subsubsection{Populating and Customising the List of Yields of Interest}
The list of yields and/or ratios of interest can be input from file using 
the function 
\texttt{InputExpYields}, provided that the file has the following format:

\small
\begin{verbatim}
             333     Exp_A    0.02    0.01
             -211    211     Exp_B  0.990   0.100
             -211    211     Exp_C  0.960   0.177
             321     -321    Exp_C  1.152   0.239
\end{verbatim}
\normalsize

\noindent
where the first line corresponds to a yield, and has format:\\

\texttt{Yield ID /t Descriptor string /t Exp. Value /t Exp. Error/n}\\

\noindent
while the remaining lines correspond to ratios, and have format:\\

\texttt{Numerator ID /t Denominator ID /t Descriptor string /t Exp. Value /t Exp. Error/n}\\

\noindent
A \texttt{TTMYield} object is created off the heap for each line in the file, with a name derived from 
the ID's and the 
descriptor. This name is determined by the private function \texttt{GetName}, which uses the base particle set to convert the particle ID's into particle names and appends the descriptor. 
In addition to all of the Monte Carlo particle ID's in the associated base particle set, the following THERMUS-defined identifiers
are also allowed: 
\begin{itemize}
\item{ID = 1: $N_{part}$,} 
\item{ID = 2: $h^-$,}
\item{ID = 3: $h^+$.}
\end{itemize} 

A \texttt{TTMYield} object can also be added to the list using \texttt{AddYield}. Such yields should, however, have names that are consistent 
with those added by the \texttt{InputExpYields} method; the \texttt{GetName} function should be used to ensure this consistency. Only 
yields with unique names can be added to the list, since it is this name which allows retrieval of the \texttt{TTMYield} objects from the 
list. If a yield with the same name already exists in the list, a warning is 
issued. The inclusion of descriptors ensures that \texttt{TTMYield} objects can always be given unique names.\\
 
\texttt{RemoveYield(Int\_t id1,Int\_t id2,TString descr)} removes from the list and deletes the 
yield with the name derived from the specified ID's and descriptor by \texttt{GetName}. The \texttt{GetYield(Int\_t id1,Int\_t id2,TString descr)} method returns the required yield.\\

\subsubsection{Generating Model Values}

Values for each of the yields of interest listed in a \texttt{TTMThermalFit} object are calculated by the function \texttt{GenerateYields}. This method uses the current parameter values and assigned 
particle sets to calculate these model values.\\

\texttt{GenerateYields} firstly calculates the 
primordial particle densities of all constituents listed in the base particle set. This it 
does by creating the relevant \texttt{TTMThermalModel} object from the base particle set and the parameters, and then calling 
\texttt{GenerateParticleDens}. In this way, the density hash table of the newly-formed \texttt{TTMThermalModel} object is 
populated with primordial densities, as well as decay contributions, according to the base particle set (recall that 
\texttt{GenerateParticleDens} automatically calculates decay contributions in addition to primordial ones). 
\texttt{GenerateYields} then iterates 
through the list of \texttt{TTMYield} objects, calculating their specific decay contributions. New model values are then inserted into these \texttt{TTMYield} objects. In addition, the total 
$\chi^2$ and quadratic deviation are calculated, based solely on the \texttt{TTMYield} objects 
which are of fit-type. \texttt{ListYields} lists all \texttt{TTMYield} objects in the list.\\

\subsubsection{Performing a Fit}

The \texttt{FitData(Int\_t flag)} method initiates a fit to all experimental yields or ratios in the \texttt{TTMYield} list which are of fit-type. With \texttt{flag=0}, a $\chi^2$ fit is performed, while \texttt{flag=1} leads 
to a quadratic deviation fit. In both cases, \texttt{fit\_function} is called. This function determines which parameters of the associated parameter set 
are to be fit, and performs the required fit using the ROOT \texttt{TMinuit} fit class. On completion, the list of \texttt{TTMYield} objects contains the model values, while the parameter set reflects the best-fit parameters. Model values are calculated by the 
\texttt{GenerateYields} method. For each \texttt{TTMYield} object in the list, a model value is calculated-- even those that have been chosen to be excluded from the actual fit. In this way, model 
predictions can be determined at the same time as a fit is performed. \texttt{ListMinuitInfo} lists 
all information relating to the \texttt{TMinuit} object, following a fit. 

\subsubsection{\texttt{TTMThermalFitBSQ}, \texttt{TTMThermalFitBQ} and \texttt{TTMThermalFitCanBSQ}}

The constructor in each of these derived classes instantiates an object with 
the specified base particle set and parameter set and inputs the 
yields listed in the specified file in the \texttt{TTMYield} list.\\

The specifics of the fit, i.e. the treatment of quantum statistics (in the 
grand-canonical ensemble and for the non-strange particles in the 
strangeness-canonical ensemble), resonance width (in all three ensembles) 
and excluded volume corrections (in the grand-canonical ensemble), are 
handled through the \texttt{SetQStats}/ \texttt{SetNonStrangeQStats}, 
\texttt{SetWidth} and 
\texttt{SetExclVol} methods, respectively. By default, both resonance width and quantum statistics are included, 
while excluded volume corrections are excluded, where applicable.\\

\subsubsection{Example}

As an example to conclude this section, consider a fit to fictitious particle ratios measured in Au+Au collisions at some 
energy. We will assume a grand-canonical ensemble, with the  
parameters $T$, $\mu_B$ and $\mu_S$ fitted, and $\mu_Q$ fixed to zero. In the grand-canonical 
ensemble, ratios are independent of the fireball radius (this is not true in the 
canonical ensemble). For this reason, there is no need to specify the treatment of the radius. 
Furthermore, we will ignore the effects of resonance width and quantum statistics.\\ 

We begin by instantiating a particle set object, based on the particle list 
distributed with THERMUS. After inputting the particle decays 
(scaled to 100\%), a parameter set is defined:

\small
\begin{verbatim}
root [ ] TTMParticleSet set("$THERMUS/particles/PartList_PPB2002.txt")
root [ ] set.InputDecays("$THERMUS/particles/",true)
root [ ] TTMParameterSetBSQ par(0.160,0.05,0.,0.,1.)
\end{verbatim}
\normalsize

\noindent
Next, we change the parameters $T$, $\mu_B$ and $\mu_S$ to fit-type, supplying sensible starting values as 
the arguments to the appropriate functions. For all other properties of the fit (step size, fit 
range etc.), we accept the default values:

\small
\begin{verbatim}
root [ ] par.FitT(0.160)
root [ ] par.FitMuB(0.05)
root [ ] par.FitMuS(0.)
\end{verbatim}
\normalsize

\noindent
Next, we prepare a file (`ExpData.txt') containing the experimental data:

\small
\begin{verbatim}

        -211    211     Exp_A   0.990   0.100
        -211    211     Exp_B   0.960   0.177
        -211    211     Exp_D   1.000   0.022
        321     -321    Exp_B   1.152   0.239
        321     -321    Exp_D   1.098   0.111
        321     -321    Exp_C   1.108   0.022
        -2212   2212    Exp_A   0.650   0.092
        -2212   2212    Exp_B   0.679   0.148
        -2212   2212    Exp_D   0.600   0.072
        -2212   2212    Exp_C   0.714   0.050
        -3122   3122    Exp_B   0.734   0.210
        -3122   3122    Exp_C   0.720   0.024
        -3312   3312    Exp_C   0.878   0.054
        -3334   3334    Exp_C   1.062   0.410
\end{verbatim}
\normalsize

\noindent
As one can see, there are multiple occurrences of the same particle--anti-particle combination. This is why 
additional descriptors are required. In this case, the descriptors list the particular experiment 
responsible for the measurement. In other situations, the descriptors may describe whether feed-down corrections 
have been employed or some other relevant detail that, together with the ID's, uniquely identifies each yield or ratio.\\   

We are now in a position to create a \texttt{TTMThermalFitBSQ} object based on the newly-instantiated 
parameter 
and particle sets and the data file. Since quantum statistics and resonance width are included by default, 
we have to explicitly turn these settings off: 

\small
\begin{verbatim}
root [ ] TTMThermalFitBSQ fit(&set,&par,"ExpData.txt")
root [ ] fit.SetQStats(kFALSE)
root [ ] fit.SetWidth(kFALSE)
\end{verbatim}
\normalsize

\noindent
Next, let us simply generate the model values corresponding to each of the \texttt{TTMYield} objects in the list, based on the current parameters. Part of the output of \texttt{ListYields} is shown here: 

\small
\begin{verbatim}
root [ ] fit.GenerateYields()
root [ ] fit.ListYields()
********************************
                                        -
                                        -
                                        -
    K+/anti-K+ Exp_B:
                        FIT YIELD
                        Experiment:    1.152     +-  0.239
                        Model:      0.979833     +-      0
                        Std.Dev.: -0.720365  Quad.Dev.: -0.175711

    K+/anti-K+ Exp_D:
                        FIT YIELD
                        Experiment:    1.098     +-  0.111
                        Model:      0.979833     +-      0
                        Std.Dev.: -1.06457  Quad.Dev.: -0.120599

    K+/anti-K+ Exp_C:
                        FIT YIELD
                        Experiment:    1.108     +-  0.022
                        Model:      0.979833     +-      0
                        Std.Dev.: -5.82578  Quad.Dev.: -0.130805

      anti-p/p Exp_A:
                        FIT YIELD
                        Experiment:     0.65     +-  0.092
                        Model:      0.535261     +-      0
                        Std.Dev.: -1.24716  Quad.Dev.: -0.21436

                                        -
                                        -
                                        -

  ******************************************************************************

\end{verbatim}
\normalsize

Finally, we perform a $\chi^2$ fit:

\small
\begin{verbatim}

root [ ] fit.FitData(0)  

                                        -
                                        -
                                        -

 COVARIANCE MATRIX CALCULATED SUCCESSFULLY
 FCN=3.54326 FROM MIGRAD    STATUS=CONVERGED     128 CALLS         129 TOTAL
                     EDM=6.43919e-06    STRATEGY= 1      ERROR MATRIX ACCURATE
  EXT PARAMETER                                   STEP         FIRST
  NO.   NAME      VALUE            ERROR          SIZE      DERIVATIVE
   1  T            1.62878e-01   1.10211e-01   2.32021e-04  -7.98809e-03
   2  muB          3.58908e-02   1.91364e-02   1.68543e-05   4.05740e-03
   3  muS          1.06828e-02   8.13945e-03   1.80744e-05   1.25485e-01
 EXTERNAL ERROR MATRIX.    NDIM=  25    NPAR=  3    ERR DEF=1
  4.593e-03  1.289e-03  5.418e-04
  1.289e-03  3.689e-04  1.548e-04
  5.418e-04  1.548e-04  6.653e-05

 PARAMETER  CORRELATION COEFFICIENTS
       NO.  GLOBAL      1      2      3
        1  0.99017   1.000  0.990  0.980
        2  0.99410   0.990  1.000  0.988
        3  0.98814   0.980  0.988  1.000
 FCN=3.54326 FROM MIGRAD    STATUS=CONVERGED     128 CALLS         129 TOTAL
                     EDM=6.43919e-06    STRATEGY= 1      ERROR MATRIX ACCURATE
  EXT PARAMETER                                  PHYSICAL LIMITS
  NO.   NAME      VALUE            ERROR       NEGATIVE      POSITIVE
   1  T            1.62878e-01   1.10211e-01   5.00000e-02   1.80000e-01
   2  muB          3.58908e-02   1.91364e-02   0.00000e+00   5.00000e-01
   3  muS          1.06828e-02   8.13945e-03   0.00000e+00   5.00000e-01
\end{verbatim}
\normalsize

Once completed, the associated parameter set contains the best-fit values for the fit parameters:

\small
\begin{verbatim}
root [ ] par.List()

 ***************************** Thermal Parameters ****************************

       T      =    0.162878     +-    0.110211      (FITTED!)
                                                     start: 0.16
                                                     range: 0.05 -- 0.18
                                                     step:  0.001

     muB      =    0.0358908    +-   0.0191364      (FITTED!)
                                                     start: 0.05
                                                     range: 0 -- 0.5
                                                     step:  0.001

     muS      =    0.0106828    +-  0.00813945      (FITTED!)
                                                     start: 0
                                                     range: 0 -- 0.5
                                                     step:  0.001

     muQ      =          0                          (FIXED)

  gammas      =          1                          (FIXED)

  radius      =          0                          (FIXED)

     muC      =          0                          (FIXED)

  gammac      =          1                          (FIXED)

                         Parameters unconstrained

  ******************************************************************************

\end{verbatim}
\normalsize


\section{Installation of THERMUS}

Having introduced the basic functionality of THERMUS in the previous section, 
we conclude by outlining the installation procedure.\\

Since several functions in THERMUS use `Numerical Recipes 
in C' code \cite{NRC} (which is under copyright), it is required that THERMUS users have their own copies 
of this software. Then, with ROOT \cite{Root} already installed, the following steps are
to be followed to install THERMUS:
\begin{itemize}
\item{Download the THERMUS source;}
\item{Set an environment variable `THERMUS' to point at the top-level directory containing the THERMUS code;}
\item{Copy the following `Numerical Recipes in C' \cite{NRC} functions to \texttt{\$(THERMUS)/nrc}:
\texttt{
\begin{tabular}{ll}
broydn.c & rsolv.c\\
fdjac.c  & fmin.c\\
lnsrch.c & nrutil.c\\
nrutil.h & qrdcmp.c\\
qrupdt.c & rotate.c\\
zbrent.c; & 
\end{tabular}
}
}
\item{Use the makefiles in \texttt{\$(THERMUS)/functions}, \texttt{\$(THERMUS)/nrc} and  
\texttt{\$(THERMUS)/main} to 
build the \texttt{libFunctions.so}, \texttt{libNRCFunctions.so} and \texttt{libTHERMUS.so} shared 
object files (run \texttt{make all} in each of these directories);}
\item{Finally, open a ROOT session, load the libraries and begin:
\small
\begin{verbatim}

root [ ] gSystem->Load("./lib/libFunctions.so");
root [ ] gSystem->Load("./lib/libNRCFunctions.so");
root [ ] gSystem->Load("./lib/libTHERMUS.so");
                          -
                          -
                          -








\end{verbatim}
\normalsize
}
\end{itemize}



\begin{thebibliography}{999}
\bibliographystyle{unsrt}


\bibitem{andronic} A. Andronic, P. Braun-Munzinger, J. Stachel, Nucl. Phys. A772 (2006) 167.

\bibitem{wheaton} J. Cleymans, H. Oeschler, K. Redlich, S. Wheaton, Phys. Rev. C73 (2006) 034905.

\bibitem{becattini} F. Becattini, J. Manninen, M. Gazdzicki, Phys. Rev. C73 (2006) 044905.

\bibitem{jaipur} For a recent review see F. Becattini, plenary talk presented at Quark Matter 2008, 
Jaipur, India, Feb 4 - 10, 2008.

\bibitem{reviewPBM}P. Braun-Munzinger, K. Redlich,  J. Stachel,
  nucl-th/0304013 and  in Quark Gluon Plasma 3, eds. R.C. Hwa and
  X.N. Wang, (World Scientific Publishing, 2004).

\bibitem{share} G. Torrieri, S. Steinke, W. Broniowski, W. Florkowski, J. Letessier, J. Rafelski, 
Comput. Phys. Commun. {\bf 167} (2005) 229.

\bibitem{sharev2} G. Torrieri, S. Jeon, J. Letessier, J. Rafelski, 
Comput. Phys. Commun. {\bf 175} (2006) 635.

\bibitem{therminator} A. Kisiel, T. Talu\'c, W. Broniowski, W. Florkowski, Comput. Phys. Commun. {\bf 174} (2006) 669.

\bibitem{Root} R. Brun and F. Rademakers, Nucl. Inst. \& Meth. in
  Phys. Res. A {\bf 389} (1997) 81.\\ 
  See also http://root.cern.ch/. 



\bibitem{LHC} J. Cleymans, I. Kraus, H. Oeschler, K. Redlich and S. Wheaton, Phys. Rev. C {\bf 74} (2006) 034903.

\bibitem{Kraus} I. Kraus, J. Cleymans, H. Oeschler, K. Redlich and S. Wheaton, J. Phys. G{\bf 32} (2006) S495.

\bibitem{Caines} H. Caines, J. Phys. G{\bf 32} (2006) S171.

\bibitem{Stiles} L.A. Stiles and M. Murray, nucl-ex/0601039.

\bibitem{Takahashi} J. Takahashi (for the STAR Collaboration), nucl-ex/0711.2273.

\bibitem{Murray} M. Murray (for the BRAHMS Collaboration), nucl-ex/0710.4576.

\bibitem{HauerViscosity} M.I. Gorenstein, M. Hauer, O.N. Moroz, nucl-th/0708.0137.

\bibitem{Witt} R. Witt, J. Phys. G{\bf 34} (2007) S921.

\bibitem{Hippo} B. Hippolyte, Eur. Phys. J. {\bf C49} (2007) 121.


\bibitem{Sahoo} J. Cleymans, R. Sahoo, D.P. Mahapatra, D.K. Srivastava and S. Wheaton, Phys. Lett. {\bf B660} (2008) 172. 


\bibitem{Hauer1} M. Hauer, V.V. Begun and M.I. Gorenstein, nucl-th/0706.3290.

\bibitem{Hauer2} M.I. Gorenstein, M. Hauer, D.O. Nikolajenko, Phys. Rev. C {\bf 76} (2007) 024901. 

\bibitem{Hauer3} V.V. Begun, M. Ga\'{z}dzicki, M.I. Gorenstein, M. Hauer, V.P. Konchakovski 
and B. Lungwitz, Phys. Rev. C {\bf 76} (2007) 024902. 



\bibitem{B1} F. Becattini, Z. Phys. C {\bf 69} (1996) 485.

\bibitem{B2} F. Becattini and U. Heinz, Z. Phys. C {\bf 76} (1997) 269.

\bibitem{C8} J. Cleymans, D. Elliott, A. Ker\"{a}nen and E. Suhonen, Phys. Rev. C {\bf 57} (1998) 3319.

\bibitem{review} K. Redlich, J. Cleymans, H. Oeschler and A. Tounsi, Acta Physica Polonica {\bf B33} (2002) 1609. 

\bibitem{heavy_ions} F. Becattini, J. Cleymans, A. Ker\"anen, E. Suhonen, K. Redlich,
Phys. Rev. C {\bf 64} (2001) 024901.

\bibitem{PBM_qd} P. Braun-Munzinger, I. Heppe, J. Stachel, 
Phys. Lett. {\bf B465} (1999) 15.

\bibitem{abundances_a} P. Braun-Munzinger \textit{et al.}, Phys. Lett. {\bf B344} (1995) 43, \textit{ibid.} {\bf B365} (1996) 1.

\bibitem{PBMRHIC} P. Braun-Munzinger, D. Magestro, K. Redlich and J. Stachel, Phys. Lett. {\bf B518} (2001) 41.

\bibitem{abundances_b} J. Sollfrank, J. Phys. G: Nucl. Part. Phys. {\bf 23} (1997) 1903.

\bibitem{Polish_a} W. Broniowski and W. Florkowski, Phys. Rev. C {\bf 65} (2002) 064905.

\bibitem{Polish_b} W. Florkowski, W. Broniowski and M. Michalec, Acta Physica Polonica {\bf B33} (2002) 761.

\bibitem{Xu} N. Xu and M. Kaneta, Nucl. Phys. {\bf A698} (2002) 306c.

\bibitem{Kaneta} M. Kaneta (for the NA44 Collaboration), J. Phys. G:
  Nucl. Part. Phys. {\bf 23} (1997) 1865;\\ 
M. Kaneta and N. Xu, J. Phys. G: Nucl. Part. Phys. {\bf 27} (2001) 589.

\bibitem{UnifiedFO} J. Cleymans and K. Redlich, Phys. Rev. Lett. {\bf 81} (1998) 5284;\\ 
J. Cleymans and K. Redlich, Phys. Rev. C {\bf 60} (1999) 054908.

\bibitem{Keranen1} A. Ker\"{a}nen and F. Becattini, nucl-th/0112021.

\bibitem{PBMCleymans} P. Braun-Munzinger, J. Cleymans, H. Oeschler and
  K. Redlich, Nucl. Phys. {\bf A697} (2002) 902.

\bibitem{RedlichTounsi}
J. Cleymans, H. Oeschler and K. Redlich, Phys. Lett. {\bf B485} (2000) 27;\\
K. Redlich, S. Hamieh and A. Tounsi, J. Phys. G: Nucl. Part. Phys. {\bf 27} (2001) 413.

\bibitem{CleymansSIS} J. Cleymans, H. Oeschler and K. Redlich, Phys. Rev. C {\bf 59} (1999) 1663.

\bibitem{SKH} J. Sollfrank, P. Koch and U. Heinz, Z. Phys. C {\bf 52} (1991) 593.

\bibitem{Kaneta1} M. Kaneta and N. Xu, nucl-th/0405068.

\bibitem{WheatonKaneta} J. Cleymans, B. K\"{a}mpfer, M. Kaneta, S. Wheaton and N. Xu, Phys. Rev. C {\bf 71} (2005) 054901.

\bibitem{BGS} F. Becattini, M. Ga\'{z}dzicki and J. Sollfrank, Eur. Phys. J. C {\bf 5} (1998) 143.

\bibitem{Raf_a} J. Rafelski, Phys. Lett. {\bf B262} (1991) 333;\\
P. Koch, B. M\"uller, J. Rafelski, Phys. Rep. {\bf 142} (1986) 167.

\bibitem{Raf_b} J. Letessier, J. Rafelski, A. Tounsi, Phys. Rev. C {\bf 50} (1994) 405;\\
C. Slotta, J. Sollfrank, U. Heinz, AIP Conf. Proc. (Woodbury) {\bf 340} (1995) 462. 

\bibitem{BecEnergyScan} F. Becattini, M. Ga\'{z}dzicki, A. Ker\"{a}nen, J. Manninen and R. Stock, Phys. Rev. C {\bf 69} (2004) 024905.

\bibitem{previous_analyzes} I.G. Bearden \textit{et al.} (NA44), Phys. Rev. C {\bf 66} (2002) 044907.

\bibitem{PRC} J. Cleymans, B. K\"ampfer and S. Wheaton, Phys. Rev. C {\bf 65} (2002) 027901, nucl-th/0110035. 

\bibitem{Nantes} J. Cleymans, B. K\"ampfer and S. Wheaton, Nucl. Phys. {\bf A715} (2003) 553c, hep-ph/0208247.

\bibitem{SQMPeter} J. Cleymans, B. K\"ampfer, P. Steinberg and S. Wheaton, J. Phys. G: Nucl. Part. Phys. {\bf 30} (2004) S595, hep-ph/0311020.

\bibitem{Rafelski99} J. Lettesier and J. Rafelski, Phys. Rev. C {\bf 59} (1999) 947.

\bibitem{VDW3} G.D. Yen, M.I. Gorenstein, W. Greiner and S.N. Yang, Phys. Rev. C {\bf 56} (1997) 2210.

\bibitem{VDW1} D.H. Rischke, M.I. Gorenstein, H. St\"ocker and W. Greiner, Z. Phys. C {\bf 51} (1991) 485.

\bibitem{VDW2} J. Cleymans, M.I. Gorenstein, J. Stalnacke and E. Suhonen, Phys. Scr. {\bf 48} (1993) 277.



\bibitem{PDG} K. Hagiwara \textit{et al.}, Phys. Rev. D {\bf 66} (2002) 010001.

\bibitem{Pythia} T. Sj\"ostrand \textit{et al.}, Comput. Phys. Commun. {\bf 135} (2001) 238.

\bibitem{perc} V. Magas, H. Satz, Eur. Phys. J. C {\bf 32} (2003) 115.

\bibitem{NRC} W. H. Press, S. A. Teukolsky, W. T. Vetterling, B. P. Flannery, Numerical Recipes in C: The Art of Scientific Computing (Cambridge University Press, Cambridge, 2002).

\bibitem{Wrob} K. Wr\`oblewski, Acta Physica Polonica {\bf B16} (1985) 379.


\end{thebibliography}
\end{document}